\def\ARXIVVERSION{1}

\documentclass[twocolumn]{autart}%

\newif\ifarxivversion
\ifdefined\ARXIVVERSION
    \arxivversiontrue
\else
    \arxivversionfalse
\fi
\newcommand{\versionedtext}[2]{\ifarxivversion #1\else #2\fi}

\usepackage{graphicx}%

\usepackage[outdir=./]{epstopdf}%

\usepackage[OT2,T1]{fontenc}
\usepackage{mleftright}%

\usepackage{amssymb}
\usepackage{mathtools}

\usepackage{bm}%

\usepackage{refcount}%

\newtheorem{theorem}{Theorem}
\newtheorem{lemma}{Lemma}
\newtheorem{proposition}{Proposition}
\newtheorem{corollary}{Corollary}
\newtheorem{assumption}{Assumption}

\newtheorem{definition}{Definition}

\newtheorem{remark}{Remark}

\mathtoolsset{showonlyrefs=true}
\allowdisplaybreaks

\DeclareMathOperator*{\argmin}{arg\,min}%

\DeclareMathOperator{\idfunc}{Id}

\newcommand{\normal}{\mathcal{N}}%

\newcommand{\comp}{\mathsf{c}}

\newcommand{\abs}[1]{| #1 |}

\DeclareMathOperator{\sat}{sat}
\DeclareMathOperator{\proj}{proj}

\begin{document}

\begin{frontmatter}

\title{Non-Asymptotic Bounds for Closed-Loop Identification of {Sub-Exponentially Growing} Nonlinear Stochastic Systems
\thanksref{footnoteinfo}}%

\thanks[footnoteinfo]{This paper was not presented at any IFAC
meeting. Corresponding author S.~Siriya.}

\author[Germany]{Seth Siriya}\ead{siriya@irt.uni-hannover.de},
\author[Australia]{Jingge Zhu}\ead{jingge.zhu@unimelb.edu.au},
\author[Australia]{Dragan Ne\v{s}i\'{c}}\ead{dnesic@unimelb.edu.au},
\author[Australia]{Ye Pu}\ead{ye.pu@unimelb.edu.au}

\address[Germany]{Leibniz University Hannover, Institute of Automatic Control, 30167 Hannover, Germany}%
\address[Australia]{University of Melbourne, Department of Electrical and Electronic Engineering, Parkville VIC 3010, Australia}%


\begin{abstract}%
We {investigate} the problem of least squares parameter estimation from single-trajectory data for discrete-time, unstable, closed-loop nonlinear stochastic systems. {Specifically, we consider nonlinear systems with linearly parametrised uncertainty and additive i.i.d. process noise, in feedback with a control policy that is intentionally perturbed by an exploratory input.}
Assuming {the open-loop dynamics satisfy a particular sub-exponential input-to-state growth property, and} a region of the state space produces informative data, we establish non-asymptotic guarantees on the estimation error at times when the state trajectory evolves in this region.
If the whole state space is informative, high-probability guarantees on the error hold for all times.
Examples are provided where our results are useful for analysis beyond existing works.
\end{abstract}

\end{frontmatter}

\section{Introduction}
System identification is the process of estimating the unknown parameters for the model of a dynamical system from measured data and is important for various applications such as time-series prediction and control.
When a parametric system model is known, least-squares (LS) estimation algorithms such as \textit{ordinary least-squares} ({OLS}) and \textit{regularised least-squares} (RLS) are popular identification strategies.
They work by choosing the estimate as the parameter minimising the average squared output/state prediction error.
Characterising the performance of these methods is important for understanding whether they will be useful for identifying a particular system.
Historically, there has been a focus on establishing asymptotic guarantees, (see \cite{lai1982least,lai1983asymptotic}), but non-asymptotic bounds on {estimation error} in linear systems have {recently} been obtained (see \cite{weyer1999finite,campi2002finite,simchowitz2018learning,faradonbeh2018finite,sarkar2019near,tsiamis2019finite,sarkar2021finite,oymak2021revisiting,dean2018regret,simchowitz2020naive,li2023non}).
Non-asymptotic results enable the study of finite-time properties like sample complexity, which is useful because only finite data can be collected before estimated parameters are used for control.
However, most real-world systems are nonlinear, motivating {their study.}

The study of non-asymptotic error bounds for nonlinear dynamical systems has been garnering increasing interest.
In \cite{foster2020learning}, error bounds are derived for a class of nonlinear systems known as \textit{generalised linear transition model} (GLTM) systems in the uncontrolled setting, assuming that the system is exponentially stable.
Systems satisfying a \textit{one-point convexity and smoothness} condition are considered in \cite{sattar2022non}, with error bounds derived for the closed-loop identification problem under a stability assumption.
In \cite{mania2020active}, LS error bounds are derived for the problem of actively identifying linearly parametrised nonlinear systems, where a computation oracle chooses controls to explore the feature space, but it is assumed that states remain bounded.
On the other hand, asymptotic consistency of RLS has recently been established for linearly parametrised nonlinear systems in \cite{liu2021convergence}, assuming that outputs are not \textit{divergent}, in the sense that the limit inferior of the output is bounded as time goes to infinity.
These works assume that the system is either stable, or trajectories are non-divergent or bounded, which is limiting when considering the controller design problem, since many controllers are specifically designed to stabilise divergent and unstable systems.
Several results instead rely on the data generated by the nonlinear system to be \textit{mixing} {such that} data measured far-apart are roughly independent.
In \cite{vidyasagar2006learning}, it is shown that \cite{weyer1999finite} can be used to establish finite-sample bounds on an identification performance criterion in parametric nonlinear systems under a mixing assumption.
Nonparametric LS estimation for dependent time-series data was considered in \cite{ziemann2022single,ziemann2022learning}, where the model is chosen from a set of functions, which is applicable to nonlinear system identification. Bounds on the in-sample prediction error were derived in these works, assuming the data are mixing.
Mixing is a useful condition since ``practically any exponentially stable system driven by i.i.d. noise with bounded variance'' is mixing \cite{karandikar2002rates}.
However, unstable linear plants, and even marginally stable ones, are not mixing.

Various results overcome limitations of previous works {to consider} unstable nonlinear systems.
In \cite{kowshik2021near}, error bounds for GLTM systems are derived assuming they are \textit{slightly unstable}.
LS error bounds are established in \cite{sattar2022finite} for \textit{marginally mean-square stable} bilinear systems.
{Although interesting}, it is desirable to address the system identification problem {more generally for} nonlinear systems with linearly parametrised uncertainty.
{A promising result in this direction is} \cite{lai1982least}, which establishes the strong consistency of estimates for general linear {regression} models assuming that the {minimum eigenvalues for the} sequence of \textit{Gramians} --- {obtained by} summing the outer product of the regressors up to a time --- grows faster than the logarithm of its maximum eigenvalue.
Another is \cite{simchowitz2018learning}, where {decreasing} non-asymptotic error bounds {can be shown} for general linear {regression} models {if the Gramians grow sub-exponentially and} the regressors satisfy a \textit{block martingale small-ball} (BMSB) condition{, related to non-degeneracy of the regressors}.
It is not {clear that linearly parametrised systems should satisfy such requirements in general.}
{Recently, \cite{musavi2024identification} {verified these conditions} for linearly parametrised dynamical systems, assuming bounded process noise, locally input-to-state stable system dynamics, and real-analytic basis functions and control policies.}
{However, investigating unstable systems with unbounded noise and non-real-analytic basis functions remains of interest.}

Motivated by these gaps, we {investigate conditions for potentially unstable nonlinear systems with linearly parametrised uncertainty that lead to decreasing} non-asymptotic estimation error bounds for RLS. We conceptually {build on} the approach in \cite{simchowitz2018learning}, but {prioritise obtaining conditions on elementary objects specific to the nonlinear system problem setup that are simple to work with}. Additionally, we aim to support the closed-loop {feedback} setting. Our contributions are as follows.

{Firstly, assuming i.i.d. sub-Gaussian noise, we show that sub-exponential growth of the Gramian can be established if the basis functions are polynomially bounded, the control policy is bounded, and the open-loop dynamics satisfy a sub-exponential input-to-state growth property. The latter property is meaningful as a systems-theoretic notion that quantitatively characterises nonlinear system instability. A decreasing error bound is then derived under an additional \emph{relaxed} \emph{persistent excitation} (PE) property, which requires that the Gramian is lower bounded by a linearly growing function over \emph{some} initial states, failure probabilities, and time intervals.}

{Secondly, we provide a condition for verifying relaxed PE called \emph{regional excitation} that only involves the basis function, control policy, and exploratory input/noise distributions.} {It} characterises the {capability} of a region {in} the state space {to produce} informative regressor data when the closed-loop trajectory evolves through it, and generalises BMSB.
{This is useful {since} nonlinear systems are not guaranteed to} produce informative data {on} the entire state space, and is exemplified in a PWA system example {satisfying} regional excitation, but {not} BMSB.
{Error bounds are then established that hold for some probabilities over a time interval where the trajectory remains inside the exciting region.}
{A numerical example applying RLS to the PWA system produces estimation errors that qualitatively match the bounds.}

Thirdly, by assuming \textit{global excitation} --- i.e., regional excitation holds over the entire state space --- high-probability error bounds that hold {for all time are obtained}.
Moreover, by further assuming that the {open-loop dynamics satisfy a polynomial input-to-state growth property}, the error bound has the convergence rate $O(\sqrt{\ln(t)/t})$.
{This matches the convergence rate {when identifying} uncontrolled linear systems with spectral radius at most 1 \cite{simchowitz2018learning} --- a special case of our more general setup.}
The usefulness of this result is showcased on an example of a {double integrator} controlled by a policy intentionally perturbed by additive exploratory {inputs}.

The work is organised as follows. In Sec.~\ref{sysid:sec:sys-id-problem}, we introduce the {considered} class of discrete-time stochastic systems and {RLS}.
Sec.~\ref{sysid:sec:estimation} contains our main results {on the sub-exponential growth of the Gramians, the decreasing error bounds under the relaxed PE property, and the investigation into regional and global excitation.}
Sec.~\ref{sysid:sec:sys-id-examples} contains the PWA system and double integrator examples demonstrating the benefit of our main results. Sec.~\ref{sysid:sec:proof-estimation} contains the proofs. Our conclusions are presented in Sec.~\ref{sysid:sec:sys-id-conclusion}.

\paragraph*{Notation}
Denote by $\mathbb{R}^{n \times m}$ the set of real matrices of dimension $n \in \mathbb{N}$ and $m \in \mathbb{N}$.
The space $\mathbb{R}^n$ stands for $\mathbb{R}^{n \times 1}$.
For a matrix $A \in \mathbb{R}^{n \times m}$, $\vert A \vert$ denotes its induced $2$-norm, $\vert A \vert_F$ denotes its Frobenius norm, and $B_r(A):=\{ \tilde{A} \in \mathbb{R}^{n \times m} \mid \vert \tilde{A} - A \vert \leq r \}$.
For a vector $v \in \mathbb{R}^n$ and a positive definite matrix $A \in \mathbb{R}^n$, $\vert v \vert_A = \sqrt{v^{\top} A v}$ denotes the weighted Euclidean norm.
Denote the unit sphere embedded in $\mathbb{R}^d$ by $\mathcal{S}^{d-1}$.
{A function $f: \mathbb{R}_{\geq 0} \rightarrow \mathbb{R}$ is said to be \emph{eventually positive} if there exists $R\geq0$ such that $f(r)>0$ for all $r\geq R$.}
Consider the function $f: \mathbb{R}_{\geq 0} \rightarrow \mathbb{R}$.
Given {an eventually positive function} $g: \mathbb{R}_{\geq 0} \rightarrow \mathbb{R}$, we say $f(r) = O(g(r))$ if ${\limsup}_{r \rightarrow \infty} \frac{{\lvert} f(r) {\rvert}}{g(r)} < \infty$, and ${f}(r) = o({g(r)})$ if $\lim_{r \rightarrow \infty} \frac{f(r)}{g(r)} = 0$.
Consider a probability space $(\Omega,\mathcal{F},\mathbb{P})$.
Consider a collection of random variables $X_1,X_2,\hdots$ taking values in $\mathcal{X}_1, \mathcal{X}_2,\hdots$ and a predicate $Q:\mathcal{X}_1 \times \mathcal{X}_2 \times \cdots \rightarrow \{ \mathrm{true}, \mathrm{false} \}$. With an abuse of notation, $\{Q(X_1,X_2,\hdots) \}$ stands for $\mleft \{ \omega \in \Omega \mid Q(X_1(\omega),X_2(\omega),\hdots) \mright \}$, and $\mathbb{P}\mleft ( Q(X_1,X_2,\hdots) \mright)$ stands for $\mathbb{P} \mleft ( \{Q(X_1,X_2,\hdots) \} \mright )$.
{Given an event $E \in \mathcal{F}$, we write ``on the event $E$, $Q(X_1,X_2,\hdots)$'', or equivalently ``$Q(X_1,X_2,\hdots)$ on the event $E$'', if $E \subseteq \{ Q(X_1,X_2,\hdots)\}$.}
{The {eventually positive} function $h: \mathbb{R}_{\geq 0} \rightarrow \mathbb{R}_{\geq 0}$ is said to be \textit{$n$-order sub-exponential} ($n$-SE) if  $\ln( h(r^n) ) = o (r)$.
The function $\alpha : \mathbb{R}_{\geq 0} \rightarrow \mathbb{R}_{\geq 0}$ is in class $\mathcal{K}$ if it is continuous, strictly increasing, and $\alpha(0) = 0$.
The function $\alpha : \mathbb{R}_{\geq 0} \rightarrow \mathbb{R}_{\geq 0}$ is in class $\mathcal{K}_{\infty}$ if $\alpha \in \mathcal{K}$ and $\lim_{s \rightarrow \infty} \alpha(s) = \infty$.
The function $\alpha : \mathbb{R}_{\geq 0} \rightarrow \mathbb{R}_{\geq 0}$ is in class $\mathcal{K}_{\infty}^{n\textnormal{-SE}}$ if $\alpha \in \mathcal{K}_{\infty}$ and $\alpha$ is $n$-SE.
A function $h:\mathbb{R}_{\geq 0} \rightarrow \mathbb{R}_{\geq 0}$ is said to be \textit{asymptotically polynomially bounded (APB)} if there exists $k \in \mathbb{N}$ such that $h(r) = O(r^k)$.}
{Let $\mathbb{I}$ denote the set of all non-empty intervals in $\mathbb{N}_0$.}
\section{Problem Setup} \label{sysid:sec:sys-id-problem}
{In this section, we formulate} the closed-loop stochastic system and estimator.

{Let $(\Omega,\mathcal{F},\mathbb{P})$ be the probability space on which all random variables are defined, and denote expectation and variance by $\mathbb{E}$ and $\mathbb{V}\mathrm{ar}$ respectively.
} Consider the discrete-time, stochastic, nonlinear system
\begin{align}
    X(t+1) = &f(X(t),U(t)) + \theta_*^{\top} \psi(X(t),U(t)) \\
    & \quad + W(t+1), \ t \in \mathbb{N}_0, \label{sysid:eqn:system-dynamics}
\end{align}
with $X(0)=x_0$. Here, {the \textit{process noise} $\{ W(t) \}_{t \in \mathbb{N}}$ is a random sequence taking values in  $\mathbb{W} {\subseteq \mathbb{R}^n}$, and}
$\{X(t)\}_{t \in \mathbb{N}_0}$ and $\{U(t)\}_{t \in \mathbb{N}_0}$ are the sequence of \textit{states} and \textit{controls} taking values in $\mathbb{X} \subseteq \mathbb{R}^n$ and $\mathbb{U} \subseteq \mathbb{R}^m$ respectively.
Moreover, $f:\mathbb{X} \times \mathbb{U} \rightarrow {\mathbb{R}^n}$ is a known \textit{nominal system model}, $\psi:\mathbb{X}\times\mathbb{U}\rightarrow \mathbb{R}^d$ is a known vector of \textit{basis functions}, $\theta_* \in \mathbb{R}^{d \times n}$ is the \textit{true unknown system parameter}, and $x_0 \in \mathbb{X}$ is the \textit{initial state}.\footnote{{{Nominal models are} useful {given} partial knowledge about the system dynamics. Consider a noisy, Euler-discretised model of voltage in an RC circuit $X(t+1) = (1 - \frac{T_d}{RC})X(t) + W(t)$, where $R,C,T_d>0$ denote resistance, capacitance, and the discretisation period, respectively. When purchasing resistors and capacitors, specifications typically provide nominal values $R_{\text{nom}},C_{\text{nom}} > 0$. This corresponds to a nominal component $f(x) = (1 - \frac{T_d}{R_{\text{nom}}C_{\text{nom}}})x$, such that $\theta_* = \frac{1}{R C} - \frac{1}{R_{\text{nom}} C_{\text{nom}}}$ and $\psi(x) = T_d x$. This can lead to improved estimation performance when $\lvert \theta_* \rvert$ is small, which will be discussed after Thm.~\ref{sysid:theorem:rls-error-exciting}. {Note that in general, the coordinates of $f(x,u)$ do not need to lie in the span of $\psi(x,u)$.}\label{fn:nominal}}}
For convenience, we also define $g(x,u,w):=f(x,u)+\theta_*^{\top}\psi(x,u)+w$, {and we only consider $g$ with the codomain $\mathbb{X}$ to ensure solutions are well-defined for all time.}

The control sequence is generated via
\begin{align}
    &U(t) = \alpha(X(t),S(t), \hat{\theta}(t-1) ), \ t \in \mathbb{N}_0. \label{sysid:eqn:controller}
\end{align}
where $\alpha: \mathbb{X} \times \mathbb{S} \times \mathbb{R}^{d \times n} \rightarrow \mathbb{U}$ is a \textit{known control policy} that sets the control input $U(t)$ according to the state $X(t)$, an injected noise term $S(t)$, and a parameter estimate $\hat{\theta}(t-1)$.
The {random} sequence {of \textit{exploratory {inputs}}} $\{S(t)\}_{t \in \mathbb{N}}$ {is sampled i.i.d. from a distribution $\mu_s$ with the support $\mathbb{S} \subseteq \mathbb{R}^q$}, and is injected to help facilitate convergence of the estimate $\hat{\theta}(t)$.
{The \textit{parameter estimates} $\{\hat{\theta}(t)\}_{t \in \mathbb{N}}$ taking values in $\mathbb{R}^{d \times n}$} are obtained via regularised least squares (RLS) estimation:\footnote{
    {We focus on RLS instead of OLS. It is more flexible in practice, allowing the incorporation of prior knowledge, and overcoming issues in OLS related to conditioning and high variance}
    {at the cost of some added bias \cite{pillonetto2022regularized}. Nevertheless, the main ideas of our analysis apply to OLS with minor modifications.} {In particular, the data-dependent error bound for RLS in Lemma~2 would be replaced by one for OLS, and upper and lower bounds would be derived for the standard Gramian $\sum_{i=1}^t Z(i) Z(i)^{\top}$ instead of $G(t) = \sum_{i=1}^t Z(i) Z(i)^{\top} + \gamma I$.}
}
\begin{align}
    \hat{\theta}(t) = \begin{cases}
        \vartheta_0, \ t  \in \{ -1,0 \}, \\
        \argmin_{\theta \in \mathbb{R}^{d \times n}} \sum_{s=1}^t \big \vert X(s) {- \theta^{\top} Z(s)} \\
        {- f(X(s-1), U(s-1)) \big \vert^2 }+ \gamma \vert \theta \vert_F^2, \ t \in \mathbb{N},
    \end{cases}\label{sysid:eqn:estimator}
\end{align}
where $\{ Z(t) \}_{t \in \mathbb{N}}$ is sequence of \textit{regressors} taking values in $\mathbb{R}^{d}$ and satisfying $Z(t) = \psi(X(t-1),U(t-1))$. {Moreover,} $\gamma > 0$ is a user-chosen regularisation parameter, and $\vartheta_0 \in \mathbb{R}^{d \times n}$ is an initial deterministic parameter estimate. Note that the RLS estimates also satisfy
\begin{align}
    &\hat{\theta}(t) = G(t)^{-1} \sum_{s=1}^t Z(s) \mleft ( X(s) \mright.  \mleft. - f(X(s-1),U(s-1)) \mright )^{\top}  \label{sysid:eqn:estimator-equivalent}
\end{align}
for $t \in \mathbb{N}$, where $G(t) = \sum_{i=1}^t Z(i)Z(i)^{\top} + \gamma I$ is the \emph{{(regularised)} Gramian}.

{Note that $\{ X(t) \}_{t \in \mathbb{N}_0}$, $\{ U(t) \}_{t \in \mathbb{N}_0}$, $\{Z(t)\}_{t \in \mathbb{N}}$ and $\{ \hat{\theta}(t) \}_{t \in \mathbb{N}}$ are defined via construction through
\eqref{sysid:eqn:system-dynamics}-\eqref{sysid:eqn:controller}, and depend on the
noise sequences $\{W(t)\}_{t \in \mathbb{N}}$ and $\{S(t)\}_{t \in \mathbb{N}_0}$.
{We assume that $f$, $\psi$, and $\alpha$, are all Borel measurable, ensuring these are all random sequences}.
}
\begin{remark}
    {We allow $S(t)$ and $\hat{\theta}(t-1)$ as inputs to our policy $\alpha$ in \eqref{sysid:eqn:controller} since this structure is common in the stochastic adaptive control literature (see \cite{guo1996self,lale2022reinforcement}).} Nevertheless, we point out that the identification results in this work are still applicable in the case of purely random controls, and the non-adaptive setting. For the former, we simply set $\alpha(x,s,\theta) = s$. For the latter, given a control policy $\alpha_2:\mathbb{X} \times \mathbb{S} \rightarrow \mathbb{U}$, we set $\alpha(x,s,\theta) = \alpha_2(x,s)$.
\end{remark}
\begin{remark}
    We {choose to} use $\hat{\theta}(t-1)$ rather than $\hat{\theta}(t)$ for the control policy in \eqref{sysid:eqn:controller} for technical reasons. Namely, this will allow us to utilise a \textit{regional excitation} condition to guarantee \textit{persistency of excitation} solely based on $\psi$, $\alpha$ and the distributions of $\{S(t)\}_{t \in \mathbb{N}_0}$ and $\{W(t)\}_{t \in \mathbb{N}}$.
\end{remark}

We recall the definition of sub-Gaussian random variables, whose distributions have tails that decay at least as fast as the Gaussian distribution.
\begin{definition} \label{sysid:def:sub-gaussian} (sub-Gaussian random variable)
    Given a sub sigma-field $\mathcal{G} \subseteq \mathcal{F}$ and scalar random variable $X$, we say $X\mid \mathcal{G}$ is $\sigma_x^2$-sub-Gaussian if for all $\lambda \in \mathbb{R}$, $\mathbb{E}[\exp(\lambda X) \mid \mathcal{G}] \leq \exp(\lambda^2 \sigma_x^2 / 2)$.
    If $X$ is an $\mathbb{R}^d$-valued random vector, we say $X\mid \mathcal{G}$ is $\sigma_x^2$-sub-Gaussian if $\zeta^{\top}X \mid \mathcal{G}$ is $\sigma_x^2$-sub-Gaussian for all $\zeta \in \mathcal{S}^{d-1}$. For both cases, we say $X$ is $\sigma_x^2$-sub-Gaussian when $\mathcal{G}$ is chosen as the trivial sigma-field $\{\emptyset, \Omega \}$.
\end{definition}

We make the following assumption on $\{W(t)\}_{t \in \mathbb{N}}$ over the probability space.

\begin{assumption} \label{sysid:assump:process-noise}
    (Independent and sub-Gaussian noise)
    \begin{enumerate}
        \item The process noise sequence $\{W(t)\}_{t \in \mathbb{N}}$ is i.i.d. with distribution denoted by $\mu_w$.
        \item The sequences $\{W(t)\}_{t \in \mathbb{N}}$ and $\{S(t)\}_{t \in \mathbb{N}_0}$ are mutually independent.
        \item $W(t)$ is zero-mean $\sigma_w^2$-sub-Gaussian for any $t \in \mathbb{N}$, i.e., $\mathbb{E}[\exp(\gamma \zeta^{\top}W(t))] \leq \exp(\gamma^2 \sigma_w^2 / 2)$ for any $\gamma \in \mathbb{R}$ and $\zeta \in \mathcal{S}^{n-1}$.
    \end{enumerate}
\end{assumption}

\begin{remark}
    The first and second conditions in Ass.~\ref{sysid:assump:process-noise} related to independence are standard in the control literature (e.g., \cite{grammatico2013discrete}). The third assumption ensures that the tails of {the distribution of $W(t)$ decay at least as fast as the tails of a Gaussian, and} is standard in the non-asymptotic system identification literature ({e.g., see} \cite{simchowitz2018learning,sarkar2019near}).
\end{remark}
\section{Main Results} \label{sysid:sec:estimation}

{This section contains the main theoretical results.
{In Sec.~\ref{sec:error-traj-excitation}, we establish sub-exponential bounds on the Gramian under suitable restrictions on the open-loop dynamics, controls, and basis functions, enabling the derivation of decreasing error bounds under an additional relaxed PE property.}
Then, in Sec.~\ref{sec:certify-excitation}, {regional and global excitation} conditions for certifying the {relaxed} PE property are investigated.}

\subsection{{Sub-Exponential Gramian and Decreasing Error Bounds}} \label{sec:error-traj-excitation}

{We begin this section by providing suitable assumptions on \eqref{sysid:eqn:system-dynamics} for sub-exponentially bounding the Gramian from above.}
Let $\phi(t,\xi,\{u(i)\}_{i = 0}^{t-1},\{w(i)\}_{i = 1}^t)$ denote the \textit{deterministic} solution at time $t \in \mathbb{N}_0$ {of the open-loop dynamics $g(x,u,w)$} given the initial state $\xi \in \mathbb{X}$, deterministic control input sequence $\{u(i)\}_{i=0}^{t-1} \in \mathbb{U}^t$ and deterministic process noise sequence $\{w(i)\}_{i = 1}^t \in \mathbb{W}^t$. We make the following {input-to-state growth assumption on $g$. It can be interpreted as a bound on} the reachable states of {the open-loop dynamics} as a function of energy-like estimates of the control inputs and process noise, {and is useful for quantifying its instability.}
\begin{assumption} \label{sysid:assump:forward-complete}
    (Sub-exponential input-to-state {growth of open-loop dynamics})
    There exist functions $\chi_1,\chi_3,\sigma_2 \in \mathcal{K}_{\infty}^{1\textnormal{-SE}}$, $\chi_4 \in \mathcal{K}_{\infty}^{2\textnormal{-SE}}$, $\chi_2,\sigma_1 \in \mathcal{K}_{\infty}$ and a constant $c_1 \geq 0$ such that $\vert \phi(t,\xi,\{u(i)\}_{i = 0}^{t-1}, \{w(i)\}_{i=1}^t) \vert \leq \chi_1(t) + \chi_2( \vert \xi  \vert )  + \chi_3 ( \sum_{i=0}^{t-1}  \sigma_1  (  \vert u(i)  \vert  )  )  + \chi_4 ( \sum_{i=1}^t  \sigma_2  (  \vert w(i)  \vert  )  ) + c_1 $
    for all $t \in {\mathbb{N}_0}$, $\xi \in \mathbb{X}$, $\{u(i)\}_{i=0}^{t-1} \in \mathbb{U}^t$ and $\{w(i)\}_{i = 1}^t \in \mathbb{W}^t$.
\end{assumption}

\begin{remark} \label{sysid:remark:forward-complete}
    {Ass.~\ref{sysid:assump:forward-complete} is well-motivated in the systems-theoretic literature, as an unrestricted version of the bound is known to be equivalent to \emph{forward-completeness} in continuous-time systems \cite{angeli1999forward}.}
    {Conveniently, a Lyapunov characterisation is possible for certifying this property.}\footnote{
        {In the (nonlinear) control literature, verifiable conditions on elementary objects from the system formulation are desirable for checking properties of the system trajectory. This is because in general, the latter is known to be hard to directly work with. Lyapunov's direct method \cite{khalil2002nonlinear} is the most popular example of this paradigm. Another example is \cite{loria2005nested}, which provided a PE condition for continuous-time systems that involves only the basis functions.}\label{fn:certificate}
    }
    {In particular, adapting the techniques from \cite{angeli1999forward} to discrete time, if there exists a function $W:\mathbb{X} \rightarrow \mathbb{R}_{\geq 0}$, $d\geq 0$, and $\alpha_1,\alpha_2,\sigma_1,\sigma_2 \in \mathcal{K}_{\infty}$ such that $\alpha_1(\lvert x \rvert) \leq W(x) \leq d + \alpha_2(\lvert x \rvert) $, and $W(g(x,u,w)) \leq 1 + \sigma_1(\lvert u \rvert) + \sigma_2(\lvert w \rvert)$ for all $x \in \mathbb{X}$, $u \in \mathbb{U}$, and $w \in \mathbb{W}$, it can be shown that bounds of the form in Ass.~\ref{sysid:assump:forward-complete} hold with $\chi_1(r) = \chi_3(r) = \chi_4(r) = \alpha_1^{-1}(5r)$, $\chi_2(r)=\alpha_1^{-1}(5\alpha_2(r))$, and $c_1 = \alpha_1^{-1}(5d)$.}
        {If this Lyapunov condition is satisfied with $\sigma_2 \in \mathcal{K}_{\infty}^{1\text{-}\textnormal{SE}}$ and $\alpha_1^{-1} \in \mathcal{K}_{\infty}^{2\text{-}SE}$, then Ass.~\ref{sysid:assump:forward-complete} holds.
    }
\end{remark}

We assume the controls generated by the policy $\alpha$ satisfy a magnitude constraint.
\begin{assumption} \label{sysid:assump:constrained-controls}
    (Constrained controls)
    There exists $u_{\textnormal{max}}\geq 0$ such that for any $x \in \mathbb{X}$, $s \in \mathbb{S}$ and $\vartheta \in \mathbb{R}^{d \times n}$, $ \vert \alpha(x,s,\vartheta) \vert \leq u_{\textnormal{max}}$.
\end{assumption}
\begin{remark} \label{sysid:remark:constrained-controls}
    Magnitude constraints on controls are natural in most real-world applications due to actuator saturation, and are satisfied in strategies like model predictive control.
    {A simple, yet illustrative controller that satisfies this is the perturbed, saturated deadbeat controller $\alpha(x,s,[\hat{a} \ \hat{b}]^{\top}) = \sat_d(-(\hat{a}/\hat{b}) x) + s$ where $x \in \mathbb{R}$, $s \in [-c,c]$ and $[\hat{a} \ \hat{b}]^{\top} \in \mathbb{R} \times \mathbb{R} \backslash \{ 0 \}$, with the constants $c,d>0$. It was combined with least-squares estimation in \cite{siriya2022learning} to obtain an adaptive control strategy that renders the states of the scalar system $X_{t+1} = a X_t + b U_t + W_t$ with $a \in {[-1,1]}$ and $b \neq 0$ mean square bounded.}
\end{remark}

We assume that the magnitude of the basis functions $\psi$ grow polynomially as a function of states and controls.

\begin{assumption} \label{sysid:assump:poly-feature-map} {
    (APB basis functions)
    There exists a {non-decreasing} APB function $\chi_5:\mathbb{R}_{\geq 0} \rightarrow \mathbb{R}_{\geq 0}$ such that $ \vert \psi(x,u)  \vert \leq \chi_5  (  \vert {[x^{\top} \ {u}^{\top}]^{\top}}  \vert  )$.}
\end{assumption}

\begin{remark}
    Ass.~\ref{sysid:assump:poly-feature-map} allows us to obtain bounds on the regressor sequence as a function of the states and controls. It is not very restrictive and allows for the selection of many common basis functions, including those with bounded magnitude and the identity mapping.
\end{remark}

{Next} we introduce a few objects required {for the Gramian upper bound}. We define the \textit{process noise bound} $\overline{w}:\mathbb{N}_0 \times (0,1) \rightarrow \mathbb{R}_{\geq 0}$, \textit{state bound} $\overline{x}: {\mathbb{N}_0} \times (0,1) \times \mathbb{X}\rightarrow \mathbb{R}_{\geq 0}$ and \textit{regressor bound} $\overline{z}:\mathbb{N} \times (0,1) \times \mathbb{X}\rightarrow \mathbb{R}_{\geq 0}$ as
\begin{align}
    &\overline{w}(t,\delta) := \begin{cases}
        \sigma_w \sqrt{{2 n \ln \mleft( \frac{{n}\pi^2t^2}{3\delta} \mright )} }, \quad t \in \mathbb{N}, \\
        0, \quad t = 0,
    \end{cases} \label{sysid:eqn:high-prob-noise-bound} \\
    &\overline{x}(t,\delta,x_0) := \chi_1(t) + \chi_2\mleft (\mleft \vert x_0 \mright \vert \mright ) + \chi_3 \mleft ( t \sigma_1 \mleft ( u_{\textnormal{max}} \mright ) \mright ) \\
    &\quad + \chi_4 \mleft ( t \sigma_2 \mleft ( \overline{w}(t,\delta) \mright ) \mright ) + c_1, \label{sysid:def:high-prob-state-bound} \\
    &\overline{z}(t,\delta,x_0) := \chi_5 \mleft (  {\sqrt{\overline{x}(t-1,\delta,x_0)^2 + u_{\textnormal{max}}^2} } \mright ), \label{sysid:def:high-prob-regressor-bound}
\end{align}
{with $\chi_1$-$\chi_4$ and $c_1$ from Ass.~\ref{sysid:assump:forward-complete},} {and $\chi_5$ from Ass.~\ref{sysid:assump:poly-feature-map}.}
Under Ass.~\ref{sysid:assump:process-noise}-\ref{sysid:assump:poly-feature-map}, these objects provide high-probability upper bounds on the magnitude of $\{W(t)\}_{t \in \mathbb{N}}$, $\{X(t)\}_{t \in \mathbb{N}_0}$ and $\{Z(t)\}_{t \in \mathbb{N}}$ respectively.

{We now arrive at our first result, on a high-probability semidefinite upper bound on the Gramian which grows sub-exponentially.}

\begin{proposition} \label{prop:sub-exp-gramian} {(Sub-exponentially bounded Gramians)}
    {Suppose Ass.~\ref{sysid:assump:process-noise}, \ref{sysid:assump:forward-complete}, \ref{sysid:assump:constrained-controls}, and \ref{sysid:assump:poly-feature-map} hold. Then, for all $x_0 \in \mathbb{X}$ and $\delta \in (0,1)$, $\mathbb{P}( \lambda_{\textnormal{max}}(G(t)) \leq \beta_{\textnormal{max}}(t,\delta,x_0), \ \forall t \in \mathbb{N}_0 ) \geq 1 - \delta$, where}
    \begin{align}
        \beta_{\textnormal{max}}(t,\delta,x_0) &:= \sum_{i=1}^t\overline{z}^2(i,\delta,x_0) + \gamma, \label{sysid:def:gram-max-bound}
    \end{align}
    {Moreover, $\beta_{\textnormal{max}}(\cdot,\delta,x_0)$ is $1$-SE.}
\end{proposition}

\begin{remark} \label{sysid:remark:relaxed-control-growth}
    {Although Ass.~\ref{sysid:assump:constrained-controls} assumes bounded controls to establish Prop.~\ref{prop:sub-exp-gramian}, alternative sufficient conditions are possible. For example, if $\alpha$ grows at most linearly in $x$, an analogous result could follow by replacing the open-loop growth condition in Ass.~\ref{sysid:assump:forward-complete} with one on the closed-loop trajectory. Specifically, the open-loop trajectory and $\{u(i)\}_{i=0}^{t-1}$ would be replaced by the closed-loop trajectory $\phi_{\textnormal{cl}}(t,\xi,\{s(i)\}_{i=0}^{t-1},\{w(i)\}_{i=1}^t)$ and $\{s(i)\}_{i=0}^{t-1}$, respectively.}
\end{remark}

{Independently, Prop.~\ref{prop:sub-exp-gramian} is useful since combined with BMSB, Thm.~2.4 in \cite{simchowitz2018learning} can be used to obtain decreasing error bounds when applied to linearly parametrised systems. As hinted at in the introduction, however, BMSB can be restrictive. This will be further explained in Sec.~\ref{sec:certify-excitation}. For now, we instead move towards an error bound result under the following \emph{relaxed} PE property.}

\begin{assumption} \label{assump:interval-pe} {(Relaxed PE)}
    {
        There exist $\mathcal{C} \subseteq (0,1) \times \mathbb{X} $, $\mathcal{T}_{\textnormal{PE}} : \mathcal{C} \rightarrow \mathbb{I}$, and $q > 0$, such that for any $(\delta,x_0) \in \mathcal{C}$,
    }
    \begin{align}
        &{\mathbb{P}\left(\lambda_{\textnormal{min}} \mleft ( G(t) \mright) \geq q(t-1) + \gamma, \forall t \in \mathcal{T}_{\textnormal{PE}}(\delta,x_0)\right) \geq 1-{\delta}.}
    \end{align}
\end{assumption}

{In the non-asymptotic literature, \emph{standard} PE requires that $\lambda_{\textnormal{min}}(G(t))$ is}
lower-bounded with high probability by a linearly growing function for all sufficiently large time, {any initial state $x_0 \in \mathbb{X}$, and any failure probability $\delta\in(0,1)$ (see \cite{tsiamis2022statistical}).}
{It} is known to be a {useful} ingredient {for obtaining decreasing error bounds for linear systems {(e.g., \cite{simchowitz2018learning} proves PE under BMSB)}, but may be too restrictive for nonlinear systems.}
{Compared to standard PE, Ass.~\ref{assump:interval-pe} is relaxed in the sense that} {the lower bound need only hold for $\delta$ and $x_0$ in some subset $\mathcal{C}\subseteq (0,1) \times \mathbb{X}$, over the time interval $\mathcal{T}_{\textnormal{PE}}(\delta,x_0)$.}

We now provide {Thm.~\ref{sysid:theorem:rls-error-exciting}} on error bounds {for RLS}.

\begin{theorem}\label{sysid:theorem:rls-error-exciting}
    (RLS error bounds under {restricted PE})
    Suppose Ass.~\ref{sysid:assump:process-noise}, \ref{sysid:assump:forward-complete}, \ref{sysid:assump:constrained-controls}, \ref{sysid:assump:poly-feature-map} and {\ref{assump:interval-pe}} are satisfied. {Then, for any {$\delta \in (0,1)$} such that $(\delta/3,x_0) \in \mathcal{C}$},
    \begin{align}
        &\mathbb{P} \mleft (\mleft \vert \hat{\theta}(t) - \theta_* \mright \vert \leq e(t,{\delta},x_0) , \ {\forall t \in \mathcal{T}_{\textnormal{PE}}({\delta/3},x_0)} \mright ) \geq 1 - {\delta}, \quad  \label{sysid:eqn:theorem-rls-error-exciting-main}
    \end{align}
    where the error bound $e(t,\delta,x_0)$ is defined as
    \begin{align}
        &e(t,\delta,x_0) := \frac{1}{ \sqrt{ { {q} (t-1) + \gamma }} } \Big( \gamma^{1/2} {\mleft \vert \theta_* \mright \vert} + \label{sysid:eqn:theorem-rls-error-exciting-e}\\
        & \ \sigma_w  \sqrt {  {8}   \mleft (  \ln \mleft ( {3} / \delta \mright )  { + n \ln (5)}      + \mleft ( d/2 \mright ) \ln \mleft (  \beta_{\textnormal{max}}\mleft ( t,{\delta/3},x_0 \mright ) \gamma^{-1}  \mright )   \mright )   }  \bigg)
    \end{align}
    {and satisfies $\lim_{t \rightarrow \infty} e(t,\delta,x_0) = 0$,}
    and $\beta_{\textnormal{max}}$ is defined in \eqref{sysid:def:gram-max-bound}.
\end{theorem}

{Since the estimation problem corresponds to linear regression, a standard self-normalised martingale inequality argument is used during the derivation of Thm.~\ref{sysid:theorem:rls-error-exciting}, producing a {data-dependent} error bound that scales with $\sqrt{\ln(\lambda_{\textnormal{max}}(G(t)))/\lambda_{\textnormal{min}}(G(t))}$ (see Lem.~\ref{sysid:lemma:data-dependent-error-bound} in Sec.~\ref{sysid:sec:proof-estimation}, or \cite{abbasi2011regret,lai1983asymptotic,simchowitz2018learning} where {similar arguments} are used).
{The decreasing bound property} {$\lim_{t \rightarrow \infty} e(t,\delta,x_0) = 0$} {then emerges from the probabilistic lower and upper bounds on the Gramian in Ass.~\ref{assump:interval-pe} and Prop.~\ref{prop:sub-exp-gramian} respectively.}
We refer the reader to Sec.~\ref{sysid:sec:proof-estimation} for the full proof.}

Thm.~\ref{sysid:theorem:rls-error-exciting} provides an upper bound $e(t,\delta,x_0)$ on the estimation error $ \vert \hat{\theta}(t) - \theta_*  \vert$ that holds uniformly over the PE interval with probability $1-{\delta}$ {if $(\delta/3,x_0) \in \mathcal{C}$}.
The importance of {$q$ from the restricted PE property in Ass.~\ref{assump:interval-pe}} is obvious when inspecting \eqref{sysid:eqn:theorem-rls-error-exciting-e}, where it can be seen that $e$ is small if the denominator dominates the numerator.
This {also} suggests the bound is smaller in systems experiencing less instability, since the numerator is smaller when {$\chi_1$--$\chi_4$ and $c_1$ from Ass.~\ref{sysid:assump:forward-complete} are smaller (through $\beta_{\textnormal{max}}(t,\delta/3,x_0)$, $\overline{z}(t,\delta/3,x_0)$, and $\overline{x}(t,\delta/3,x_0)$).}
Moreover, {for a fixed $t$,} the bound {can be made smaller by taking} $\gamma$ {sufficiently} large and $\lvert \theta_* \rvert$ {sufficiently small}, motivating the selection of a good nominal model $f$, as discussed earlier.\footnotemark[\getrefnumber{fn:nominal}]

\begin{remark} \label{sysid:remark:error-bound-other-intervals} {
    With some additional work on top of Thm.~\ref{sysid:theorem:rls-error-exciting}, high-probability bounds that hold over all $t \in \mathbb{N}$ can be derived, rather than just for $t \in \mathcal{T}_{\textnormal{PE}}(\delta{/3},x_0)$.
    {
        This is obtained by replacing $q(t-1)$ in $e(t,\delta,x_0)$ with $0$ for $t < \min \mathcal{T}_{\textnormal{PE}}(\delta{/3},x_0)$, and with $q(T_{\textnormal{excited}}(\delta{/3},x_0)-1)$ for $t > \max \mathcal{T}_{\textnormal{PE}}(\delta{/3},x_0)$.
    }
    However, {these additional bounds before and after the interval $\mathcal{T}_{\textnormal{PE}}(\delta{/3},x_0)$} are monotonically increasing in $t$ {since} the influence of $\beta_{\textnormal{max}}(t,\delta/3,x_0)$ {in the numerator is not offset by the linear growth in the denominator of the bound}, in contrast to $t \in \mathcal{T}_{\textnormal{PE}}(\delta{/3},x_0)$ {where} the {denominator grows}{, resulting in bounds which can} decrease within the PE interval, and {not} afterwards. This qualitatively {matches} the behaviour of the system in Example~1 in Sec.~\ref{sysid:sec:sys-id-example-pwa-behaviour}, where the estimation error stops improving after {some} time.}

\end{remark}

\subsection{{Certifying Relaxed PE}} \label{sec:certify-excitation}

{{Decreasing} error bounds were obtained in Thm.~\ref{sysid:theorem:rls-error-exciting} under {the relaxed PE property from} Ass.~\ref{assump:interval-pe}. However, this is not straightforward to check{;} it involves the Gramian, which in turn depends on system trajectories.
{It is preferable to instead have checkable certificate conditions.}\footnotemark[\getrefnumber{fn:certificate}]
In {this section, we explore such a condition called \emph{regional excitation}. It is introduced and shown to verify Ass.~\ref{assump:interval-pe} in Sec.~\ref{sec:regional-excitation}.}
The case where the condition holds globally is investigated in Sec.~\ref{sec:global-excitation}.
}

\subsubsection{{Regional Excitation}} \label{sec:regional-excitation}

{Firstly, we introduce some required notation.}
For any Borel measurable function $h:\mathbb{S} \times \mathbb{W} \rightarrow \mathbb{R}$, we denote $\mathbf{E} [ h(S,W)  ] := \int_{\mathbb{W}} \int_{\mathbb{S}} h(s',w') d\mu_s(s') d\mu_w(w')$ and $\mathbf{V}\mathrm{ar}(h(S,W)):=\mathbf{E} [  ( h(S,W) - \mathbf{E} [ h(S,W) ]  )^2  ]$.
{Moreover, given a predicate $Q:\mathbb{S}\times \mathbb{W} \rightarrow \{ \mathrm{true},\mathrm{false} \}$, we define $\mathbf{P}(Q(S,W)):=\mathbf{E} [ \mathbf{1}_{\{ (s,w) \in \mathbb{S} \times \mathbb{W} \mid Q(s,w) \}}(S,W) ]$.}

We now define regional and global excitation as follows.

\begin{definition} \label{sysid:def:excitation}
    (Regional {and global} excitation)
    The feature map $\psi$ and controller family $\alpha$ are said to be \emph{regionally excited} with noise distribution $\mu_s$ and $\mu_w$ over $\mathcal{X} \subseteq \mathbb{X}$ with
    {finite constants $c_{\textnormal{PE}},p_{\textnormal{PE}} > 0$} if {$\mathcal{X}\oplus\mathbb{W}\subseteq\mathbb{X}$ and} for all $\vartheta \in \mathbb{R}^{d \times n}$, $x \in \mathcal{X}$ and $\zeta \in \mathcal{S}^{d-1}$,
    \begin{align}
        {\mathbf{P}\mleft (  \mleft \vert \zeta^{\top} \psi \mleft ( x + W , \alpha(x+W,S,\vartheta) \mright ) \mright \vert^2 \geq c_{\textnormal{PE}} \mright ) \geq p_{\textnormal{PE}}.} \label{sysid:eqn:excitation}
    \end{align}
    For convenience, we say $(\psi,\alpha,\mu_s,\mu_w)$ is {\emph{$(\mathcal{X},c_{\textnormal{PE}},p_{\textnormal{PE}})$-regionally excited}}.
    {Moreover, we say $(\psi,\alpha,\mu_s,\mu_w)$ is \emph{$(c_{\textnormal{PE}},p_{\textnormal{PE}})$-globally excited} if $(\psi,\alpha,\mu_s,\mu_w)$ is $(\mathbb{X},c_{\textnormal{PE}},p_{\textnormal{PE}})$-regionally excited.}
\end{definition}

Intuitively, {regional excitation holds} if for any state $x$ inside {$\mathcal{X}$} and parameter $\vartheta \in \mathbb{R}^{d \times n}$, when we corrupt the state by a process noise $W \sim \mu_w$, sample the exploratory {input} $S \sim \mu_s$, compute the controls based on the corrupted state and exploratory {input} $U = \alpha(x + W,S,\vartheta)$, {and} apply the basis function to the corrupted state and controls {to obtain $\psi(x+W,\alpha(x+W,S,\vartheta))$, the distribution of this random variable is non-degenerate in every direction of the regressor space.}
{This allows a linearly growing lower bound on $G(t)$ to be established for trajectories evolving in $\mathcal{X}$, as required in Ass.~\ref{assump:interval-pe}, such that the region $\mathcal{X}$ can be viewed as \textit{informative}} for estimation.
{Global excitation in Def.~\ref{sysid:def:excitation} is a strengthening of regional excitation so that it holds over the entire state space $\mathbb{X}$. Such a condition may be restrictive {in general}, but allows us to make stronger statements on the estimation error bound $e$.}

{Compared to Ass.~\ref{assump:interval-pe}, which involves the Gramian, and hence the trajectories of the system in \eqref{sysid:eqn:system-dynamics}, the regional excitation condition in Def.~\ref{sysid:def:excitation} is easier to work with since it is} characterised entirely by the basis functions $\psi$, control policy $\alpha$, and the distribution of the process {noise} and exploratory {input} $\mu_w$, $\mu_s$.
{Notably, it also does not require $\theta_*$ to be known, which is useful given the goal of learning $\theta_*$.}
{Later, in Sec.~\ref{sysid:sec:sys-id-example-integrators}, we will illustrate the usefulness of this condition in a closed-loop scenario where a stochastic double integrator is controlled by a policy consisting of an arbitrary bounded function of the state and parameter estimate, plus a bounded i.i.d. noise.
}

Moreover, there are nonlinear systems where regional excitation holds, but other conditions such as BMSB from \cite{simchowitz2018learning} {are not applicable (although the two are related, see Rem.~\ref{remark:regional-excitation-bmsb}), and hence neither are works that rely on such conditions (e.g., \cite{li2023non,musavi2024identification})}.
{This can happen when there are regions of the state-input space where the basis function is constant, and the system can enter these regions. A PWA system example demonstrating this is provided later on in Sec.~\ref{sysid:sec:sys-id-example-pwa}.}
{{Another relevant scenario is when the basis functions} approach a constant value {in the limit (e.g., $\psi(x)=\tanh(x)$)} and the trajectories are able to grow unboundedly due to the noise and underlying system dynamics.}
{The error bound results in this paper will handle such situations, as long as the assumptions to sub-exponentially bound the Gramian also hold (see Prop.~\ref{prop:sub-exp-gramian}).}

\begin{remark} \label{remark:regional-excitation-bmsb}
    {{Regional excitation} in Def.~\ref{sysid:def:excitation} actually implies a restricted version of BMSB (see Def.~\ref{sysid:def:bmsb}, or \cite{simchowitz2018learning}) with block size $k=1$, in the sense that for all $\zeta \in \mathcal{S}^{d-1}$ and ${j \geq 1}$, $\mathbb{P}( \lvert \zeta^{\top} Z(j+1) \rvert \geq {\sqrt{ {c_{\textnormal{PE}}} }} \mid \{ X(m) \}_{m=0}^{j{-1}}, \{ U(m) \}_{m=0}^{{j-1}} ) \geq p_{\textnormal{PE}}$ holds on the event $\{ \mathbb{E}[ X(j) \mid X(j-1),U(j-1) ]\in\mathcal{X} \}$. In the global case where $\mathcal{X} = \mathbb{X}$, this recovers standard BMSB.
    Rather than considering a general BMSB-like formulation in Def.~\ref{sysid:def:excitation} --- which would allow relaxation of Ass.~\ref{sysid:assump:process-noise} to handle noise from a martingale difference sequence --- we opt for the current problem-specific notion since it is simpler to interpret. Note that an analogous condition to Def.~\ref{sysid:def:excitation} that considers block sizes $k>1$ can also be formulated, but is hard to verify in general since it involves several iterations of the nonlinear function $g$, and therefore also the unknown parameter $\theta_*$. Thus, we stick to the $k=1$ case for simplicity. }
\end{remark}

{{Convenient ways to verify regional excitation exist}, such as Lem.~\ref{sysid:lemma:regional-excitation-sufficient}, which involves bounding the {mean and variance} of $\mleft \vert \zeta^{\top} \psi ( x + W , \alpha(x+W,S,\vartheta) \mright ) {\vert} $ from below and above respectively.}
{We refer to \versionedtext{Sec.~\ref{sysid:sec:proof-regional-excitation-moments}}{the Supplementary Results in \cite{siriya2024non}} for its proof.}
\begin{lemma} \label{sysid:lemma:regional-excitation-sufficient}
    {Consider a side{vector of basis functions} $\psi$, a control policy $\alpha$, and noise distributions $\mu_w,\mu_s$ satisfying Ass.~\ref{sysid:assump:process-noise}. Suppose there exists a subset $\mathcal{X} \subseteq \mathbb{X}$ and constants $c_{\textnormal{PE1}}>0$ and $c_{\textnormal{PE2}}\geq 0$ such that for all $\vartheta \in \mathbb{R}^{d \times n}$, $x \in \mathcal{X}$ and $\zeta \in \mathcal{S}^{d-1}$, $\mathbf{E} [  \vert \zeta^{\top} \psi  ( x + W , \alpha(x+W,S,\vartheta)  )  \vert ]  \geq c_{\textnormal{PE1}}$ and $\mathbf{V}\mathrm{ar} (  \vert \zeta^{\top} \psi  ( x + W , \alpha(x+W,S,\vartheta)  )  \vert  ) \leq c_{\textnormal{PE2}}$.
    Then, $(\psi,\alpha,\mu_s,\mu_w)$ is $(\mathcal{X},c_{\textnormal{PE}},p_{\textnormal{PE}})$\textit{-regionally excited} with
    $p_{\textnormal{PE}} := \frac{1}{4} ( \frac{ c_{\textnormal{PE}2}}{c_{\textnormal{PE}1}^2} + 1 )^{-1}$ and $c_{\textnormal{PE}} := \frac{1}{4} c_{\textnormal{PE}1}^2.$}
\end{lemma}
{In general, applying Lem.~\ref{sysid:lemma:regional-excitation-sufficient} and verifying Def.~\ref{sysid:def:excitation} may be {challenging}, {because} it involves verification over all $\mathcal{X}$, alongside the entire parameter space $\mathbb{R}^{d \times n}$ {since $\alpha$ is parameter-dependent}. These difficulties can be addressed in certain scenarios. For example, boundedness properties of the policy can be utilised to satisfy Lem.~\ref{sysid:lemma:regional-excitation-sufficient}, as shown later on in Sec.~\ref{sysid:sec:sys-id-example-integrators}. Moreover, if $\alpha(x,s,\vartheta) = \alpha_2(x,s,\proj_{\Theta}(\vartheta))$ {for some policy $\alpha_2:\mathbb{X}\times\mathbb{S}\times\Theta \rightarrow {\mathbb{U}}$} {and} $\mathcal{X} \times \Theta$ is compact, Lem.~\ref{sysid:lemma:regional-excitation-sufficient} can be empirically verified by discretising $\mathcal{X} \times \Theta \times \mathcal{S}^{d-1}$ {at some points} and applying Monte Carlo approximation.}

{We assume that regional excitation holds in this section.}

\begin{assumption} \label{sysid:assump:regional-excitation}
    $(\psi,\alpha,\mu_s,\mu_w)$ is {$(\mathcal{X}_{\textnormal{PE}},c_{\textnormal{PE}},p_{\textnormal{PE}})$\textit{-regionally excited}} for some set $\mathcal{X}_{\textnormal{PE}} \subseteq \mathbb{X}$ and constants {$c_{\textnormal{PE}},p_{\textnormal{PE}} > 0$}.
\end{assumption}

{In order to connect Ass.~\ref{sysid:assump:regional-excitation} with Ass.~\ref{assump:interval-pe}, some objects need to be introduced.} We first define the \textit{{one-step predicted} reachable set map} as
\begin{align}
    \Gamma(\mathcal{X}) := \mleft \{ g(x,u,0) \mid x \in \mathcal{X}, \mleft \vert u \mright \vert \leq u_{\textnormal{max}} \mright \}, \ \mathcal{X} \subseteq \mathbb{X},
\end{align}
which is the {set of} reachable states from $\mathcal{X} \subseteq \mathbb{X}$ {when the} control input {magnitude is bounded by} $u_{\textnormal{max}}$ and the process noise {is set to zero}.

Next, we define the \textit{excited time bound} as
\begin{align}
    &T_{\textnormal{excited}}(\delta,x_0) := \\
    & \quad \sup \mleft \{ T \in \mathbb{N} \mid \Gamma \mleft ( B_{\overline{x}(T-1,{\delta/{2}},x_0)}(0) {\cap \mathbb{X}} \mright )  \subseteq \mathcal{X}_{\textnormal{PE}} \mright \}, \quad \label{sysid:def:T-excited}
\end{align}
$T_{\textnormal{excited}}(\delta,x_0)$ can be interpreted as a high-probability lower bound on the horizon over which the predicted state $g(X(t-1),U(t-1),0)$ remains inside $\mathcal{X}_{\textnormal{PE}}$.
\begin{remark}
    {
        The centring of the ball in the definition of $T_{\textnormal{excited}}$ in \eqref{sysid:def:T-excited} is without loss of generality. To consider a ball centred at another point $x^0 \in \mathcal{X}_{\textnormal{PE}}$, one could instead apply the state transformation $\tilde{X}(t) = X(t) - x^0$, yielding the state-transformed system $\tilde{X}(t+1) =  f(\tilde{X}(t) + x^0,U(t)) + \theta_*^{\top}\psi(\tilde{X}(t) + x^0,U(t)) + W(t+1) - x^0$. A ball around zero for this system corresponds to a ball around $x^0$ for the original system.
    }
\end{remark}
Moreover, define the \textit{burn-in time bound} as
\begin{align}
        &T_{\textnormal{burn-in}}(\delta,x_0) := \inf \bigg \{  T \in \mathbb{N} \;\biggm\vert\; t \geq \frac{2}{(1 - \ln(2))  p_{\textnormal{PE}}}  \\
        & \quad \times \Bigg ( d \ln \mleft (1 + \frac{16 \sum_{i=1}^t \overline{z}^2 \mleft (i, {\delta/{2}}, x_0 \mright ) }{ c_{\textnormal{PE}} p_{\textnormal{PE}} (t-1) } \mright ) \\
        & \quad + \ln \mleft ( \frac{\pi^2 (t - T + 1)^2 }{{{3}\delta}} \mright ) \Bigg )  + 1 \text{ for all } t \geq T {\geq 2} \bigg \}, \label{sysid:eqn:theorem-rls-error-exciting-T-burn-in}
\end{align}

Intuitively, $T_{\textnormal{burn-in}}(\delta,x_0)$ can be viewed as a high-probability upper bound on the time it takes for PE to start holding, assuming that the predicted state $g(X(t-1),U(t-1),0)$ remains inside $\mathcal{X}_{\textnormal{PE}}$.

{We now provide Prop.~\ref{prop:regional-excitation-pe}, which establishes that the regional excitation condition in Def.~\ref{sysid:def:excitation} certifies excitation in the sense of Ass.~\ref{assump:interval-pe}.}

\begin{proposition} \label{prop:regional-excitation-pe}
    {
        Suppose Ass.~\ref{sysid:assump:process-noise}, \ref{sysid:assump:forward-complete}, \ref{sysid:assump:constrained-controls}, \ref{sysid:assump:poly-feature-map}
        {and \ref{sysid:assump:regional-excitation} are satisfied.}
        Then, Ass.~\ref{assump:interval-pe} is satisfied with $\mathcal{C}:= \{ ({\delta,x_0}) \in {(0,1)\times\mathbb{X}} \mid T_{\textnormal{burn-in}}({\delta,x_0}) \leq T_{\textnormal{excited}}({\delta,x_0}) \}$, $q = c_{\textnormal{PE}}p_{\textnormal{PE}}/4$, and $\mathcal{T}_{\textnormal{PE}}(\delta,x_0):=\{T_{\textnormal{burn-in}}(\delta,x_0), \hdots,$ $T_{\textnormal{excited}}(\delta,x_0) \}$, with ${T}_{\textnormal{burn-in}}({\delta,x_0})$, ${T}_{\textnormal{excited}}({\delta,x_0})$ from \eqref{sysid:eqn:theorem-rls-error-exciting-T-burn-in}, {\eqref{sysid:def:T-excited}} respectively. Furthermore, $T_{\textnormal{burn-in}}(\delta,x_0) < \infty$ for all $\delta \in (0,1)$ and $x_0 \in \mathbb{X}$. {It follows that the premise of Thm.~\ref{sysid:theorem:rls-error-exciting} is also satisfied.}
    }
\end{proposition}

The intuition here is that {at times $t$ where $T_{\textnormal{burn-in}}({\delta,x_0}) \leq t \leq T_{\textnormal{excited}}({\delta,x_0})$ with $(\delta,x_0)\in \mathcal{C}$}, {$G(t)$ is well-conditioned} and $g(X(t-1),U(t-1),0)$ remains inside {the informative region} $\mathcal{X}_{\textnormal{PE}}$. {This allows $G(t)$ to be bounded from below by a linearly growing function, establishing relaxed PE in Ass.~\ref{assump:interval-pe}}.
{Thm.~\ref{sysid:theorem:rls-error-exciting} can then be used to obtain estimation error bounds over this time interval. In contrast, the bounds in \cite{simchowitz2018learning} hold over all $\delta \in (0,1)$, $x_0 \in \mathbb{X}$, and $t \geq T_{\textnormal{burn-in}}(\delta,x_0)$ due to the stricter BMSB requirement.}

{A central question} {is whether the condition $T_{\textnormal{burn-in}}({\delta,x_0}) \leq T_{\textnormal{excited}}({\delta,x_0})$ is satisfied for some $\delta$ and $x_0$, such that $\mathcal{C}\neq\emptyset$.}
{For a given system, this can be directly checked by constructing $T_{\textnormal{burn-in}}(\delta,x_0)$ and $T_{\textnormal{excited}}(\delta,x_0)$ according to \eqref{sysid:eqn:theorem-rls-error-exciting-T-burn-in} and \eqref{sysid:def:T-excited}.}
However, qualitatively, it is more likely to hold in systems with a larger excitation region, a greater degree of excitation, and with less instability. This is because $T_{\textnormal{burn-in}}(\delta,x_0)$ is smaller {for slower growing} $\overline{z}(t,\delta/{2},x_0)$ and larger ${c_{\textnormal{PE}}}$ and $p_{\textnormal{PE}}$, and $T_{\textnormal{excited}}({\delta},x_0)$ is larger for larger $\mathcal{X}_{\textnormal{PE}}$ and slower {growing} ${\overline{x}}(t,\delta/{2},x_0)$.

\subsubsection{{Global Excitation}} \label{sec:global-excitation}

We now suppose that global excitation holds and establish Cor.~\ref{sysid:corollary:rls-error-global} under this assumption.
\begin{assumption} \label{sysid:assump:global-excitation}
    $(\psi,\alpha,\mu_s,\mu_w)$ is {$(c_{\textnormal{PE}},p_{\textnormal{PE}})$\textit{-globally excited}} for some constants {$c_{\textnormal{PE}},p_{\textnormal{PE}} > 0$}.
\end{assumption}

{
    In general, Ass.~\ref{sysid:assump:global-excitation} is harder to verify than Ass.~\ref{sysid:assump:regional-excitation}.
    It requires verification over the entire state space $\mathbb{X}$, which may be unbounded.
    However, it can be checked in specific scenarios.
    The simplest case is when $\psi$ consists of the state and control coordinates themselves (e.g., see the controlled-double-integrator example in Sec.~\ref{sysid:sec:sys-id-example-integrators}).
    Another case is when global excitation can be reduced to regional excitation.
    For example, when $\psi$ is periodic in the state, global verification can be reduced to verification over a bounded domain, as with Fourier basis functions.
    Another situation arises when there is a bounded set $\mathcal{K}$ such that $\psi$ is regionally excited outside $\mathcal{K} \oplus \mathbb{W}$, where $\mathbb{W}$ is bounded.
    In this case, it remains only to verify regional excitation over $\mathcal{K}\oplus\mathbb{W}$.
}

\begin{corollary} (RLS error bounds under global excitation) \label{sysid:corollary:rls-error-global}
    Suppose Ass.~\ref{sysid:assump:process-noise}, \ref{sysid:assump:forward-complete}, \ref{sysid:assump:constrained-controls}, \ref{sysid:assump:poly-feature-map} and \ref{sysid:assump:global-excitation} are satisfied. Then, for any $x_0 \in \mathbb{X}$ and $\delta \in (0,1)$,
    \begin{align}
        &\mathbb{P} \mleft (\mleft \vert \hat{\theta}(t) - \theta_* \mright \vert \leq e(t,\delta,x_0) , \ \forall t \geq T_{\textnormal{burn-in}}({\delta}/{3},x_0) \mright ) \\
        &\geq 1 - {\delta}, \label{eqn:error-bound-global-excitaiton}
    \end{align}
    with $\lim_{t \rightarrow \infty} e(t,\delta,x_0) = 0$.
\end{corollary}
Cor.~\ref{sysid:corollary:rls-error-global} can be viewed as an extension of Thm.~\ref{sysid:theorem:rls-error-exciting} {
    combined with Prop.~\ref{prop:regional-excitation-pe},
} to consider the case where global excitation holds.
{This allows one to take $x_0 \in \mathbb{X}$ and $\delta \in (0,1)$ arbitrarily, resulting in the same type of \textit{high-probability bounds} observed for {linear regression under BMSB and} linear systems \cite{simchowitz2018learning,sarkar2019near}. This is because $T_{\textnormal{excited}}(\delta,x_0) = \infty$ for all $(\delta,x_0)\in (0,1) \times \mathbb{X}$, since} the exciting region is now the entire state space $\mathbb{X}$, and the system never escapes this region.

{Although $e$ {decreases} towards zero, we cannot say anything about {the convergence rate yet}. Strengthening Ass.~\ref{sysid:assump:forward-complete} to Ass.~\ref{sysid:assump:poly-forward-complete} {allows us to} upper bound {the rate}, which is provided in Cor.~\ref{sysid:corollary:rls-error-global-poly}.}
\begin{assumption}{(Polynomial input-to-state {growth of open-loop dynamics}) \label{sysid:assump:poly-forward-complete}
    Ass.~\ref{sysid:assump:forward-complete} holds with $\chi_1, \chi_3, \chi_4, \sigma_2$ as APB functions.}
\end{assumption}
\begin{corollary} \label{sysid:corollary:rls-error-global-poly}
    {Suppose Ass.~\ref{sysid:assump:process-noise}, \ref{sysid:assump:constrained-controls}, \ref{sysid:assump:poly-feature-map}, \ref{sysid:assump:global-excitation} and \ref{sysid:assump:poly-forward-complete} are satisfied. Then, {the results of Cor.~\ref{sysid:corollary:rls-error-global} hold with} $e(t,\delta,x_0) = O(\sqrt{\ln(t)/t})$ {for any $\delta \in (0,1)$ and $x_0 \in \mathbb{X}$}.}
\end{corollary}
{The convergence rate in Cor.~\ref{sysid:corollary:rls-error-global-poly} {matches that for} linear systems {taking} the form $X(t+1)=AX(t)+W(t)$ with $W(t) \overset{\mathrm{i.i.d.}}{\sim} \mathcal{N}(0,\sigma_w^2 I)$ and spectral radius at most 1 \cite{simchowitz2018learning}, {which is a special case of our more general setup.}}
{For this class of systems, our bound can also be compared in terms of other relevant parameters, like the dimensionality $n$ of $A$, and the controllability Gramian $\Gamma_k = \sum_{i=0}^{k-1} A^i (A^i)^{\top}$. In particular, \cite{simchowitz2018learning} shows that the error is upper bounded by $\tilde{O}(\sqrt{n/(t \lambda_{\textnormal{min}}(\Gamma_k))})$ for any $k$, whereas on this system our bound is $\tilde{O}(\sqrt{n/t})$, with $\tilde{O}(\cdot)$ hiding logarithmic factors.
The term $\Gamma_k$ does not appear in our bound since this involves a $k$-step BMSB argument, but our analysis corresponds to a $1$-step analysis (recall Rem.~\ref{remark:regional-excitation-bmsb}).}
{This gap can be closed by extending regional excitation to cover $k>1$ steps, but we stick to $k=1$ for simplicity, as explained in Rem.~\ref{remark:regional-excitation-bmsb}.}
\section{Examples} \label{sysid:sec:sys-id-examples}

In this section, we provide several examples demonstrating the usefulness of our results.
In Sec.~\ref{sysid:sec:sys-id-example-pwa}, we consider a PWA system example. We first show that it satisfies {Ass.~\ref{sysid:assump:process-noise}--\ref{sysid:assump:poly-feature-map} for bounding the Gramian sub-exponentially}.
Then, in Sec.~\ref{sysid:sec:sys-id-example-pwa-benefit}, we show
{that the BMSB condition requirement is too strong for the system, motivating the use of regional excitation from Ass.~\ref{sysid:def:excitation} (implying a generalisation of BMSB, cf. Rem.~\ref{remark:regional-excitation-bmsb}), before demonstrating in Sec.~\ref{sysid:sec:sys-id-example-pwa-behaviour} that our bound qualitatively captures the behaviour of the system.}
{Subsequently, in Sec.~\ref{sysid:sec:sys-id-example-integrators}, we derive non-asymptotic error bounds for RLS identification of a stochastic double integrator --- an open-loop unstable system --- under the influence of {an arbitrary bounded policy with additive input noise}, and characterise its convergence rate, making use of Cor.~\ref{sysid:corollary:rls-error-global-poly}.}

\subsection{Example 1: Piecewise Affine System {with Random Controls}} \label{sysid:sec:sys-id-example-pwa}

{In Example~1, we aim to show the benefit of the regional excitation condition compared to the BMSB condition, and that the bond in Thm.~\ref{sysid:theorem:rls-error-exciting} qualitatively makes sense. To this end,}
we consider discrete-time PWA stochastic system dynamics of the same form as \eqref{sysid:eqn:system-dynamics}, with
\begin{align}
    f(x,u) = 1, \  \psi(x,u) = \begin{bmatrix}
        x & &
        \mathbf{1}_{\{  \tilde{x} \mid \tilde{x} \leq \overline{x} \}}(x) \cdot u
    \end{bmatrix}^{\top} \label{sysid:eqn:sys-example-pwa-dynamics}
\end{align}
and $\theta_* = [
        1 \
        0.1
    ]^{\top}$, {$\bar{x}>0$}.
Moreover, the controls are chosen as i.i.d. noise, such that the policy $\alpha$ \eqref{sysid:eqn:controller} has the form
\begin{align}
    \alpha(x,s,\vartheta) = s, \label{sysid:eqn:sys-example-pwa-controller}
\end{align}
and the distribution $\mu_s$ of the injected noise term is $\mathrm{Uniform}([-\overline{u},\overline{u}])$ with $\overline{u} > 0$.
The process noise $\{W(t)\}_{t \in {\mathbb{N}}}$ is chosen to satisfy Ass.~\ref{sysid:assump:process-noise}, where $\mu_w$ is selected as the {normal} distribution $\mathcal{N}(0,\sigma_w^2)$ with $\sigma_w > 0$.

We now verify {Ass.~\ref{sysid:assump:forward-complete}--\ref{sysid:assump:poly-feature-map}}.
To verify Ass.~\ref{sysid:assump:forward-complete}, note that for all $t \in \mathbb{N}_{{0}}$, $\xi \in \mathbb{X}$, $\{u(i)\}_{i=0}^{t-1} \in \mathbb{U}^t$ and $\{w(i)\}_{i = 1}^t \in \mathbb{W}^t$, the trajectory bound $\mleft \vert \phi(t,\xi, \{ u(i) \}_{i=0}^{t-1}, \{ w(i) \}_{i=1}^t ) \mright \vert \leq t + \mleft \vert \xi \mright \vert  + 0.1 \mleft ( \sum_{i=0}^{t-1} \mleft \vert u(i) \mright \vert \mright ) + \sum_{i=1}^t \mleft \vert w(i) \mright \vert$
holds. Thus, ${\chi_1(\cdot) =} \chi_2(\cdot) = \sigma_1(\cdot) = \chi_4(\cdot) = \sigma_2(\cdot) = \idfunc$, $\chi_3 = 0.1 \idfunc (\cdot)$, and $c_1 = 0$. The functions here {are} all in $\mathcal{K}_{\infty}^{2\textnormal{-SE}}$, thus verifying Ass.~\ref{sysid:assump:forward-complete}. Ass.~\ref{sysid:assump:constrained-controls} is satisfied with $u_{\textnormal{max}} = \overline{u}$ since the control policy is simply chosen to be the injected noise, whose support is $[-\overline{u},\overline{u}]$. To verify Ass.~\ref{sysid:assump:poly-feature-map}, note that $\mleft \vert \psi(x,u) \mright \vert = \mleft \vert (x, {\mathbf{1}_{\{ \tilde{x} \mid \tilde{x} \leq \overline{x} \}}(x)} \cdot u ) \mright \vert \leq \mleft \vert {[x^{\top} \ {u}]^{\top}} \mright \vert$,
such that Ass.~\ref{sysid:assump:poly-feature-map} is satisfied with $\chi_5 (\cdot)
= \idfunc(\cdot)$.

\subsubsection{The Benefit of Regional Excitation} \label{sysid:sec:sys-id-example-pwa-benefit}
{The difficulty of parameter estimation in PWA systems has previously been recognised in \cite{mania2020active}.}
In this section, {we show that the BMSB condition} --- commonly used to establish PE in some linear and linearly parametrised systems
(e.g., \cite{sattar2022finite,li2023non})
--- {does not hold} for the {introduced} PWA system under natural considerations. However, {we show that} our regional excitation condition (i.e., Ass.~\ref{sysid:assump:regional-excitation}) is satisfied.

Firstly, we recall the definition of the BMSB condition.
\begin{definition} \label{sysid:def:bmsb}
(Martingale Small-Ball \cite[Def. 2.1]{simchowitz2018learning}) Given $k \in \mathbb{N}$, $\Gamma_{\text{sb}} \succ 0$ and $p {\in (0,1]}$, an $\mathbb{R}^d$-valued random process $\{Z(t)\}_{t \in \mathbb{N}}$ adapted to a filtration $\{\mathcal{F}(t)\}_{t \in \mathbb{N}_0}$ is said to satisfy the $(k,\Gamma_{\text{sb}},p)$-block martingale small-ball (BMSB) condition, if, for any $\zeta \in \mathcal{S}^{d-1}$ and $j \geq 0$, $\frac{1}{k} \sum_{i=1}^k \mathbb{P} (\abs{\zeta^{\top}Z(j+i)} \geq \sqrt{\zeta^{\top} \Gamma_{\text{sb}} \zeta} \mid \mathcal{F}(j)) \geq p$
a.s.
\end{definition}
{We establish in Prop.~\ref{sysid:prop:bmsb-failure} that starting from any initial state $x_0 \in \mathbb{R}$, the regressors are unable to satisfy the BMSB condition for our PWA system.}
\begin{proposition} \label{sysid:prop:bmsb-failure}
    Consider the PWA system from \eqref{sysid:eqn:sys-example-pwa-dynamics}, \eqref{sysid:eqn:sys-example-pwa-controller}, {with the initial state} satisfying $x_0 \in \mathbb{R}$. Let $\mathcal{F}(t) = \sigma \mleft ( \{Z(i)\}_{i \leq t} \mright )$ for $t \in \mathbb{N}$, {with $\mathcal{F}(0)$ defined as the trivial sigma-field.}
    Then, there does not exist $k \in \mathbb{N}$, $\Gamma_{\textnormal{sb}} \succ 0$, and $p {\in(0,1]}$, such that $\{Z(t)\}_{t \in \mathbb{N}}$ adapted to $\{\mathcal{F}(t)\}_{t \in \mathbb{N}_0}$ satisfies the $(k,\Gamma_{\textnormal{sb}},p)$-BMSB condition.
\end{proposition}
\sloppy
{Prop.~\ref{sysid:prop:bmsb-failure} holds because regardless of the chosen $k,\Gamma_{\textnormal{sb}},p$, we can always find $j$ and $\zeta$ to establish that $\mathbb{P}  ( \frac{1}{k} \sum_{i=1}^k \mathbb{P}  (  \vert \zeta^{\top}Z(j+i)  \vert \geq \sqrt{ \zeta^{\top} \Gamma_{\textnormal{sb}} \zeta } \mid \mathcal{F}(j)  ) < p   ) > 0$. This is because of the unbounded noise, which causes a non-zero probability that $X(j)$ will be far above $\overline{x}$, such that for any $i \in \{1,\hdots, k\}$, there is a low conditional probability that $X(j+i) < \overline{x}$ will hold, and hence a low conditional probability (smaller than $p$) that the regressor projected onto $\zeta = {[0 \ 1]^{\top}}$ will satisfy $\vert \zeta^{\top}Z(j+i) \vert = \vert U(t+1) \vert = \vert \mathbf{1}_{\{ X(t+1) \leq \overline{x}\}}S(t) \vert \geq \sqrt{(0,1)^{\top}\Gamma_{\textnormal{sb}}(0,1)}$.}

{Although Prop.~\ref{sysid:prop:bmsb-failure} shows that the BMSB condition fails for any initial state $x_0$, the following result shows that our regional excitation condition can still be verified on the PWA system over a certain subset of the state space. }
\begin{proposition} \label{sysid:prop:pwa-regional-excitation-success}
    Consider $(\psi,\alpha,\mu_s,\mu_w)$ from the PWA system \eqref{sysid:eqn:sys-example-pwa-dynamics}, \eqref{sysid:eqn:sys-example-pwa-controller}.     $(\psi,\alpha,\mu_s,\mu_w)$ is $(\mathcal{X}_{\textnormal{PE}},{c_{\textnormal{PE}}},{p_{\textnormal{PE}}})$\textit{-regionally excited}, with {$\mathcal{X}_{\textnormal{PE}}:= (-\infty,0.9\overline{x}]$, $c_{\textnormal{PE}} := \frac{1}{4} c_{\textnormal{PE}1}^2$ and $p_{\textnormal{PE}} := \frac{1}{4} ( \frac{ c_{\textnormal{PE}2}}{c_{\textnormal{PE}1}^2} + 1 )^{-1}$, with the constants $c_{\textnormal{PE1}}$, $c_{\textnormal{PE2}}$ defined as}
    $
        c_{\textnormal{PE1}}:= \frac{b_w b_s}{\sqrt {b_w^2 + b_s^2 }}$, and  $c_{\textnormal{PE2}}:=\max(\sigma_w^2,\frac{1}{3}\overline{u}^2),
    $
    and the constants $b_w,b_s$ defined as
    $b_w := \sigma_w {\frac{\varphi(0) - \varphi(0.1\bar{x}/\sigma_w)}{  \Phi(0.1\bar{x}/\sigma_w) - \Phi(0)}} ( \Phi ( \frac{0.1 \overline{x}}{\sigma_w}  ) - \Phi ( \frac{-0.1 \overline{x}}{\sigma_w}  )  ) $,
        and $
        b_s := \frac{1}{2}\overline{u}  ( \Phi ( \frac{0.1 \overline{x}}{\sigma_w}  ) - \Phi ( \frac{-0.1 \overline{x}}{\sigma_w}  )  ).
    $
    Here, $\varphi(\cdot)$ and $\Phi(\cdot)$ denote the probability density and cumulative distribution functions of the standard normal distribution, respectively.
\end{proposition}
{The key reason regional excitation can be verified in Prop.~\ref{sysid:prop:pwa-regional-excitation-success} is that regional excitation in Def.~\ref{sysid:def:excitation} only requires the conditions in \eqref{sysid:eqn:excitation} to be evaluated over the region $\mathcal{X}_{\textnormal{PE}} \subseteq \mathbb{X}$ where the generated data is informative.
{This implies a restricted notion of the BMSB condition holds on a subset of the sample space (cf. Rem.~\ref{remark:regional-excitation-bmsb}). In contrast, standard BMSB has an \textit{almost surely} requirement that cannot be fulfilled,} since the unbounded noise causes the states to enter a non-informative region of the state space with non-zero probability. }

Since the regional excitation condition is established in Prop.~\ref{sysid:prop:pwa-regional-excitation-success}, our main results from Sec.~\ref{sec:regional-excitation} can be applied to bound the estimation error of the PWA system. This highlights the advantage of our analysis over those relying on the BMSB condition for establishing PE.

\subsubsection{{Simulation of the PWA System}} \label{sysid:sec:sys-id-example-pwa-behaviour}

\begin{figure}[ht]
    \centering
    \includegraphics[width=0.45\textwidth]{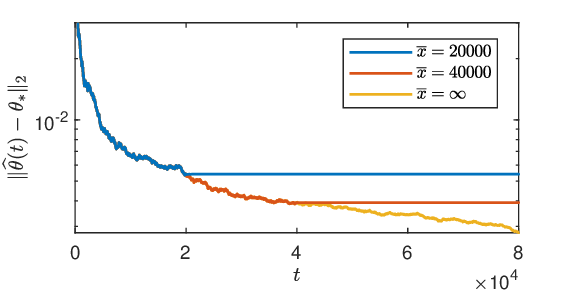}
    \caption{Log scale plot of estimation error averaged over 100 sample paths}
    \label{sysid:fig:example-1-averaged-errors}
\end{figure}

{Recall that the bounds in Thm.~\ref{sysid:theorem:rls-error-exciting} may decrease over a finite interval, but stops decreasing thereafter, as noted in Remark~\ref{sysid:remark:error-bound-other-intervals}. In this section, we will simulate the PWA system }
with varying $\overline{x}$ and observe the estimation errors over time,
{and show that simulations exhibit the same qualitative behaviour as the theoretical bounds.}
In particular, we simulated the PWA system for $\overline{x}\in \{{20000},{40000},\infty\}$, $\overline{u}={0.2}$, and ${\sigma_w}=0.1$, with $\gamma=0.0001$ and $x_0 = {0}$. The RLS estimation errors averaged over 100 sample paths are shown in Figure~\ref{sysid:fig:example-1-averaged-errors}.

{When} $\overline{x} = {20000}$ {or} $\overline{x}={40000}$, the average estimation error {appears} to improve over a finite interval {before levelling off, with this occurring earlier for $\overline{x}=20000$.} {When} $\overline{x}= \infty$, the average error continues improving throughout the simulation. This qualitatively {agrees with} Thm.~\ref{sysid:theorem:rls-error-exciting} and {Rem.}~\ref{sysid:remark:error-bound-other-intervals}{, which noted that the bound may decrease over the interval $T_{\textnormal{burn-in}}(\delta/3,x_0)\leq t \leq T_{\textnormal{excited}}(\delta/3,x_0)$, whereas its extension beyond this interval does not decrease.} {For} $\delta = 0.4$, the {computed} burn-in time was $T_{\textnormal{burn-in}}(\delta/3,x_0)={8476}$ {for all three values of $\overline{x}$, while the corresponding excited times were} ${10651}$ for $\overline{x}={20000}$, ${21052}$ for $\overline{x}={40000}$, and $\infty$ for $\overline{x}=\infty$. This trend is consistent with the observation that larger $\overline{x}$ leads to a larger time interval over which the error is decreasing in Figure~\ref{sysid:fig:example-1-averaged-errors}. {Comparison with} Figure~\ref{sysid:fig:example-1-averaged-errors} {nevertheless shows} that these quantities are conservative{. In more unstable systems, it is also possible that $T_{\textnormal{excited} }(\delta/3,x_0) < T_{\textnormal{burn-in}}(\delta/3,x_0)$, in which case $\mathcal{T}_{\textnormal{PE}}=\emptyset$.}

\subsection{Example 2: Controlled Double Integrator} \label{sysid:sec:sys-id-example-integrators}

{In Example~2, we show how Cor.~\ref{sysid:corollary:rls-error-global-poly} can be used to establish non-asymptotic error bounds for the closed-loop identification problem with a policy which {is not necessarily} stabilising. {Such results can be useful for proving stability of adaptive control strategies that rely on the convergence of parameter estimates, like \cite{siriya2022learning,siriya2023stability}.} We consider a stochastic double integrator:}
\begin{align}
    \underbrace{\begin{bmatrix}
        X_1(t+1) \\ X_2(t+1)
    \end{bmatrix}}_{X(t+1)} = \begin{bmatrix}
        1 & 0\\ 1 & 1
    \end{bmatrix} \underbrace{\begin{bmatrix}
        X_1(t) \\ X_2(t)
    \end{bmatrix}}_{X(t)} + \begin{bmatrix}
        1 \\ 0
    \end{bmatrix} U(t) + W(t{+1}).
\end{align}
The nominal system $f$, unknown parameter $\theta_*$, and basis function $\psi$ in \eqref{sysid:eqn:system-dynamics} correspond to
$f(x,u) = 0$,
$\theta_* = \begin{bmatrix}
        1 & 0 & 1 \\ 1 & 1 & 0
    \end{bmatrix}^{\top}$, and
$\psi(x,u) = \begin{bmatrix}
    x_1 & x_2 & u
\end{bmatrix}^{\top}$.
Moreover, we consider scalar controls $U(t)$, and our policy $\alpha$ in \eqref{sysid:eqn:controller} is chosen as the sum of a scalar parameter-dependent state-feedback control policy $\alpha_2$ and a scalar i.i.d. noise: $U(t) = \alpha(X(t),S(t),\hat{\theta}(t-1)) = \alpha_2(X(t),\hat{\theta}(t-1)) +S(t).$
We allow $\alpha_2$ to be an arbitrary control policy that is Borel measurable and satisfies $\mleft \vert \alpha_2(x,\vartheta) \mright \vert \leq \overline{u}_1$ with $\overline{u}_1 > 0$. We emphasise that $\alpha_2$ \textit{does not} need to be a stabilising policy. We choose the distribution $\mu_s$ of the injected noise term as $\mathrm{Uniform}([-\overline{u}_2,\overline{u}_2])$ with $\overline{u}_2 > 0$.
The process noise $\{W(t)\}_{t \in {\mathbb{N}}}$ taking values in $\mathbb{R}^2$ is sampled from the {normal} distribution $\mathcal{N}(0,\sigma_w^2 I)$ with $\sigma_w > 0$, which satisfies Ass.~\ref{sysid:assump:process-noise}.
Moreover, Ass.~\ref{sysid:assump:constrained-controls} is satisfied with $u_{\textnormal{max}}=\overline{u}_1 + \overline{u}_2$ and Ass.~\ref{sysid:assump:poly-feature-map} is satisfied with $\chi_5(\cdot) = \mathrm{Id}(\cdot)$.

{We now establish that Ass.~\ref{sysid:assump:poly-forward-complete} is satisfied. Recall that
$\phi(t, \xi, \{ u(i) \}_{i=0}^{t-1},\{ w(i) \}_{i=1}^t )$
is the deterministic trajectory starting at
$\xi = (\xi_1,\xi_2) \in \mathbb{R}^2$
under controls
$ \{ u(i) \}_{i=0}^{t-1} \in \mathbb{R}^t$
and disturbances
$\{ w(i) \}_{i=1}^t = \{ (w_1(i),w_2(i) {)} \}_{i=1}^t \in \mathbb{R}^{2t}$ for $t \in \mathbb{N}_0$.
We denote
$(x_1(t),x_2(t)) = \phi(t, \xi, \{ u(i) \}_{i=0}^{t-1},\{ w(i) \}_{i=1}^t )$.
Then,
$\vert x_1(t) \vert =  \vert \xi_1 + \sum_{i = 0}^{t-1}u(i) + \sum_{i=1}^tw_{{1}}(i)  \vert \leq \vert \xi_1 \vert + \sum_{i = 0}^{t-1} \vert u(i) \vert + \sum_{i=1}^t \vert w_1(i) \vert$,
and $\vert x_2(t) \vert =  \vert \xi_2 + \sum_{i=0}^{t-1} x_1(i) + \sum_{i=1}^t w_2(i)  \vert
    \leq \vert \xi_2 \vert + \vert \xi_1 \vert t + (t-1) \sum_{i=0}^{t-2} \vert u(i) \vert + (t-1) \sum_{i=1}^{t-1} \vert w_1(i) \vert  + \sum_{i=1}^t \vert w_2(i) \vert$,
so
\begin{align}
    &\vert x(t) \vert \leq \vert x_1(t) \vert + \vert x_2(t) \vert
    \leq  5 \vert \xi \vert^2  + \frac{5}{2}  \mleft (\sum_{i = 0}^{t-1} \vert u(i) \vert  \mright )^2 \\
    & \qquad \qquad + 5  \mleft(\sum_{i=1}^t \vert w(i) \vert  \mright)^2 + \frac{5}{2} t^2 + \frac{5}{2} \label{sysid:eqn:sys-id-example-integrators-3}
\end{align}
where \eqref{sysid:eqn:sys-id-example-integrators-3} follows from $\vert x \vert_1\leq \sqrt{n} \vert x \vert $ for all $x \in \mathbb{R}^n$, the AM-GM inequality, and the QM-AM inequality.
Thus, Ass.~\ref{sysid:assump:poly-forward-complete} is satisfied with $\chi_{{2}}(\cdot) = \chi_4(\cdot) = 5 \idfunc(\cdot)^2$, $\chi_{{1}}(\cdot) = \chi_3(\cdot)  = \frac{5}{2} \idfunc(\cdot)^2$, {$\sigma_1(\cdot)=\sigma_2(\cdot)=\idfunc(\cdot)$} and $c_1 = \frac{5}{2}$.}

{We now verify Ass.~\ref{sysid:assump:global-excitation}. Suppose $x \in \mathbb{R}^2$, $\vartheta \in \mathbb{R}^{3 \times 2}$, and $\zeta = {[\zeta_1^{\top} \ \zeta_2]^{\top}} \in \mathcal{S}^{2}$ (where $\zeta_1 \in \mathbb{R}^2$ and $\zeta_2 \in \mathbb{R}$). Then,
\begin{align}
    &\mathbf{E} \mleft [ \mleft \vert \zeta^{\top} \psi(x+W, \alpha(x+W,S,\vartheta) \mright \vert \mright ] \\
    & \geq
    \mathbf{E} \mleft [ \mleft \vert \zeta_1^{\top} x + \zeta_2 \mathbf{E} \mleft [ \alpha_2(x+W,\vartheta) \mright ] + \zeta_2 S \mright \vert\mright ] \label{sysid:eqn:sys-id-example-integrators-5} \\
    &\geq \vert \zeta_2 \vert \mathbf{E}\mleft[ \mleft \vert S \mright \vert \mright] = \vert \zeta_2 \vert \frac{\overline{u}_2}{2} \label{sysid:eqn:sys-id-example-integrators-6}
\end{align}
where \eqref{sysid:eqn:sys-id-example-integrators-5} follows from the tower property and the independence of $W$ and $S$, and \eqref{sysid:eqn:sys-id-example-integrators-6} follows from the optimality property of medians and $ \mathbf{E}\mleft[ \mleft \vert S \mright \vert \mright] = \frac{\overline{u}_2}{2} $. Similarly,
\begin{align}
    &\mathbf{E} \mleft [ \mleft \vert \zeta^{\top} \psi(x+W, \alpha(x+W,S,\vartheta) \mright \vert \mright ] \\
    & \geq \mathbf{E} \mleft [ {\mleft \vert \zeta_1^{\top}W \mright \vert} \mright ] - \mleft \vert \zeta_2 \mright \vert \mathbf{E} \mleft [ \mleft \vert \alpha_2(x+W,\vartheta) \mright \vert\mright ] \label{sysid:eqn:sys-id-example-integrators-9} \\
    & {\geq \vert \zeta_1 \vert \sigma_w \sqrt{\frac{2}{\pi}} - \vert \zeta_2 \vert \overline{u}_1} \label{sysid:eqn:sys-id-example-integrators-11}
\end{align}
where \eqref{sysid:eqn:sys-id-example-integrators-9} follows from the tower property, the independence of $W$ and $S$, the reverse triangle inequality, and {the optimality property of medians, and} \eqref{sysid:eqn:sys-id-example-integrators-11} follows from the fact that $\alpha_2$ is upper bounded by $\overline{u}_1$, and
{$ \mathbf{E}[  \vert \zeta_1^{\top} W \vert  ] \geq \vert \zeta_1 \vert \sigma_w \sqrt{\frac{2}{\pi}}$ since $\zeta_1^{\top} W\sim \normal(0, \vert \zeta_1 \vert^2 \sigma_w^2)$.}
Note that if $\vert \zeta_2 \vert \leq {\frac{\sigma_w\sqrt{2/\pi}}{2\mleft(\overline{u}_1+\sigma_w\sqrt{2/\pi}\mright)}}$, then $\vert \zeta_1 \vert \geq 1 - {\frac{\sigma_w\sqrt{2/\pi}}{2\mleft(\overline{u}_1+\sigma_w\sqrt{2/\pi}\mright)}}$, and so
{$\vert \zeta_1 \vert \sigma_w \sqrt{ \frac{2}{\pi} } -  \vert \zeta_2  \vert \overline{u}_1 \geq \mleft(1-\frac{\sigma_w\sqrt{2/\pi}}{2\mleft(\overline{u}_1+\sigma_w\sqrt{2/\pi}\mright)}\mright)\sigma_w\sqrt{\frac{2}{\pi}}-\frac{\sigma_w\sqrt{2/\pi}\,\overline{u}_1}{2\mleft(\overline{u}_1+\sigma_w\sqrt{2/\pi}\mright)}=\frac{\sigma_w}{\sqrt{2\pi}}$.}
Moreover, if $\vert \zeta_2 \vert > {\frac{\sigma_w\sqrt{2/\pi}}{2\mleft(\overline{u}_1+\sigma_w\sqrt{2/\pi}\mright)}}$, then {$\vert \zeta_2 \vert \frac{\overline{u}_2}{2} > \frac{\sigma_w\sqrt{2/\pi}\,\overline{u}_2}{4\mleft(\overline{u}_1+\sigma_w\sqrt{2/\pi}\mright)}$}. Combining this with
\eqref{sysid:eqn:sys-id-example-integrators-6} and \eqref{sysid:eqn:sys-id-example-integrators-11}, we have
$
    \mathbf{E}  [  \vert \zeta_1^{\top}(x+W) + \zeta_2 (\alpha_2(x+W,\vartheta) + S)  \vert  ]
    \geq {\min\mleft(\frac{\sigma_w}{\sqrt{2\pi}},\frac{\sigma_w\sqrt{2/\pi}\,\overline{u}_2}{4\mleft(\overline{u}_1+\sigma_w\sqrt{2/\pi}\mright)}\mright)}.
$
On the other hand, we have
\begin{align}
    &\mathbf{V}\mathrm{ar} \mleft ( \mleft \vert \zeta^{\top} \psi(x+W, \alpha_2(x+W,\vartheta) + S {)} \mright \vert \mright ) \\
    &\leq 3 \mleft (\mathbf{E} \mleft [ \mleft ( \zeta_1^{\top} W \mright )^2 \mright ] + \mathbf{E} \mleft [ \mleft( \zeta_2 S \mright )^2 \mright ] \mright. \\
    & \quad + \mleft. {\mathbf{E} \mleft [ \mleft ( \zeta_2 \mleft (\alpha_2(x+W,\vartheta) - \mathbf{E} \mleft[ \alpha_2(x+W,\vartheta) \mright ] \mright ) \mright )^2 \mright ] } \mright ) \quad \label{sysid:eqn:sys-id-example-integrators-13} \\
    &\leq 3  \max \mleft ( \sigma_w^2, \frac{\overline{u}_2^2}{3} + {4\overline{u}_1^2} \mright ) , \label{sysid:eqn:sys-id-example-integrators-14}
\end{align}
where \eqref{sysid:eqn:sys-id-example-integrators-13} follows from the QM-AM inequality, and \eqref{sysid:eqn:sys-id-example-integrators-14} follows from $\mathbf{E}  [ WW^{\top}  ]  \leq \sigma_w^2 I$, $\mathbf{E} [ S^2  ] \leq \frac{\overline{u}_2^2}{3}$ and \\
$\mathbf{E}  [  ( \zeta_2  (\alpha_2(x+W,\vartheta) - \mathbf{E} [ \alpha_2(x+W,\vartheta)  ]  )  )^2  ] \leq 4 \bar{u}_1^2$.
Thus, we find that {$(\psi,\alpha,\mu_s,\mu_w)$ is $(c_{\textnormal{PE}},p_{\textnormal{PE}})$\textit{-globally excited} with
$c_{\textnormal{PE}} := \frac{1}{4} c_{\textnormal{PE}1}^2$ and $p_{\textnormal{PE}} := \frac{1}{4} ( \frac{ c_{\textnormal{PE}2}}{c_{\textnormal{PE}1}^2} + 1 )^{-1}$, where
$c_{\textnormal{PE1}}:={\min\mleft(\frac{\sigma_w}{\sqrt{2\pi}},\frac{\sigma_w\sqrt{2/\pi}\,\overline{u}_2}{4\mleft(\overline{u}_1+\sigma_w\sqrt{2/\pi}\mright)}\mright)}$ and $c_{\textnormal{PE2}}:=3  ( \max  ( \sigma_w^2, \frac{\overline{u}_2^2}{3}  ) + 4\overline{u}_1^2  ) $}, satisfying Ass.~\ref{sysid:assump:global-excitation}.}

Since Ass.~\ref{sysid:assump:process-noise}, \ref{sysid:assump:constrained-controls}, \ref{sysid:assump:poly-feature-map}, \ref{sysid:assump:global-excitation} and \ref{sysid:assump:poly-forward-complete} are all satisfied, we can use Cor.~\ref{sysid:corollary:rls-error-global-poly} to conclude that for all $x_0 \in \mathbb{R}^2$ and $\delta \in (0,1)$, \\$\mathbb{P}  ( \vert \hat{\theta}(t) - \theta_*  \vert \leq e(t,\delta,x_0) , \ \forall t \geq T_{\textnormal{burn-in}}(\delta/3,x_0)  ) \geq 1 - \delta$ where $e(t,\delta,x_0) = O  ( \sqrt{\ln(t)/t}  )$.
{Note that this bound has not yet been established in the existing literature for the system considered here.} In particular, \cite{li2023non} derive convergence rates for non-asymptotic error bounds in closed-loop linear system identification with {arbitrary bounded policies and additive noise}, but assume the trajectory is uniformly bounded over time, which is not satisfied by this example.
{Although the theory in \cite{simchowitz2018learning} is applicable, steps like this example are required to show BMSB.}
\section{Proofs} \label{sysid:sec:proof-estimation}

{We now provide {proofs of the} results in this work.
Sec.~\ref{sysid:sec:proofs-error-bounds-regional} contains the {proofs for Sec.~\ref{sysid:theorem:rls-error-exciting}}, and Sec.~\ref{sec:proof-certify-excitation} contains the {proofs for Sec.~\ref{sec:certify-excitation}}.}

\subsection{{Proofs for Sub-Exponential Gramian and Decreasing Error Bound}} \label{sysid:sec:proofs-error-bounds-regional}
{In this section, we provide {supporting lemmas for} {Prop.~\ref{prop:sub-exp-gramian} and Thm.~\ref{sysid:theorem:rls-error-exciting}}, and the {proofs of Prop.~\ref{prop:sub-exp-gramian} and Thm.~\ref{sysid:theorem:rls-error-exciting}}.} {The proofs of supporting lemmas are deferred to Sec.~\ref{sysid:sec:proof-supporting-results}.}

{{In order to prove Prop.~\ref{prop:sub-exp-gramian},} we need an upper bound on $\lambda_{\textnormal{max}}(G(t))$. Since $G(t) = \sum_{i=1}^t Z(i) Z(i)^{\top} + \gamma I$.
{To do this, we first derive} high-probability upper bounds on $\vert Z(t) \vert$ and $\vert X(t) \vert$ uniformly over all time.}
\begin{lemma} \label{sysid:lemma:high-prob-state-bound}
    \sloppy
    (High-Probability State and Regressor Bounds)
    Suppose Ass.~\ref{sysid:assump:process-noise}, \ref{sysid:assump:forward-complete}, \ref{sysid:assump:constrained-controls} and \ref{sysid:assump:poly-feature-map} hold. Then, for any $x_0 \in \mathbb{X}$ and $\delta \in (0,1)$,
    $
        \mathbb{P}  (  \vert X(t-1)  \vert \leq \overline{x}(t-1,\delta,x_0)
        \ \text{and} \  \vert Z(t)  \vert \leq  \overline{z}(t,\delta,x_0) , \ \forall t \in \mathbb{N}  ) \geq 1 - \delta,
    $
    with $\overline{x}$ and $\overline{z}$ defined in \eqref{sysid:def:high-prob-state-bound} and \eqref{sysid:def:high-prob-regressor-bound} respectively.
\end{lemma}

{We are now ready to prove Prop.~\ref{prop:sub-exp-gramian}.}

\begin{pf*}{Proof of Prop.~\ref{prop:sub-exp-gramian}}
    Suppose $\delta \in (0,1)$ and $x_0 \in \mathbb{X}$. {Then, $\mathbb{P}( \lambda_{\textnormal{max}}(G(t)) \leq \beta_{\textnormal{max}}(t,\delta,x_0), \ \forall t \in \mathbb{N}_0 ) \geq 1 - \delta$ follows immediately from Lem.~\ref{sysid:lemma:high-prob-state-bound} and \eqref{sysid:def:gram-max-bound}.}

    {We now show that $\beta_{\textnormal{max}}(\cdot,\delta,x_0)$ is $1$-SE. For all $t\in\mathbb{N}$,}
    $
        {\beta_{\textnormal{max}}(t,\delta,x_0)}
        {=\sum_{i=1}^t\left(\overline z^2(i,\delta,x_0)+\frac{\gamma}{t}\right)}
        {\leq t\left(\overline z^2(t,\delta,x_0)+\frac{\gamma}{t}\right),}
    $
    {where the inequality holds since $i \mapsto \overline{z}(i,\delta,x_0)$ is non-decreasing as $\chi_1$--$\chi_5$, $\sigma_2$, and $\overline{w}$ are non-decreasing.} Moreover, from Ass.~\ref{sysid:assump:poly-feature-map}, there exist {$\tilde{a}_1>0$ and} $\tilde{a}_2, \tilde{a}_3 \geq 0 $ such that $\chi_5(r) \leq \tilde{a}_1 r^{\tilde{a}_2} + \tilde{a}_3$ for $r \geq 0$, and so
    \begin{align}
        & {\overline{z}}  (t, \delta, x_0  ) =  \chi_5  (  \vert  (  \chi_1(t{-1}) + \chi_2 ( \vert x_0  \vert  )  \\
        & \quad + \chi_3  ( {(t-1)} \sigma_1  ( u_{\textnormal{max}}  )  ) \\
        & \quad + \chi_4  ( {(t-1)} \sigma_2  ( \overline{w}({t-1},\delta)  )  ) + c_1, u_{\textnormal{max}}  )  \vert  ) \\
        &\leq  \tilde{a}_1  (  \chi_1(t) + \chi_2 ( \vert x_0  \vert  ) + \chi_3  ( t \sigma_1  ( u_{\textnormal{max}}  )  ) \\
        &\quad + \chi_4  ( t \sigma_2  ( \overline{w}(t,\delta)  )  ) + c_1 + u_{\textnormal{max}}  )^{ \tilde{a}_2} + \tilde{a}_3. \label{sysid:eqn:lemma-sub-exponential-gramian-3}
    \end{align}
    {Therefore,}
    \begin{align}
         & {\ln \left( \beta_{\textnormal{max}}(t,\delta,x_0) \right) } \leq {\ln (t) + 2 \ln  (\tilde{a}_1  (  \chi_1(t) + \chi_2 ( \vert x_0  \vert  )  }\\
        &{+ \chi_3  ( t \sigma_1  ( u_{\textnormal{max}}  ) ) + \chi_4  ( t \sigma_2  ( \overline{w}(t,\delta)  )  ) + c_1 + u_{\textnormal{max}}  )^{\tilde{a}_2}} \\
        &  {+ \tilde{a}_3 + \sqrt{\gamma}  ).}\label{sysid:eqn:lemma-sub-exponential-gramian-1}
    \end{align}
    Additionally, since $\sigma_2\in \mathcal{K}_{\infty}$ and $\sqrt{( \cdot )} \in \mathcal{K}_{\infty}$, then $\sigma_2  ( \sigma_w \sqrt{(\cdot)}  )\in \mathcal{K}_{\infty}$, so using the weak triangle inequality, we have $\sigma_2  ( \overline{w}  (t,\delta  )  ) = \sigma_2  ( \sigma_w \sqrt{ {n (} 2 \ln ( \frac{ {n} \pi^2}{3\delta} ) + 4 \ln  ( t  ) {)}  }  )
    \leq \sigma_2  ( \sigma_w \sqrt{ 4 {n} \ln ( \frac{{n}\pi^2}{3\delta} ) }  ) + \sigma_2  ( \sigma_w \sqrt{ 8 {n} \ln  ( t  )   }  )$.
    It is easy to see that $\sigma_w \sqrt{ 8 {n} \ln  ( t  )   } = o(\ln (t))$. Then, from
    {\versionedtext{Lem.~\ref{sysid:lemma:sub-exp-equiv}}{\cite[Lem.~11]{siriya2024non}}},
    and the fact that $\sigma_2 \in \mathcal{K}_{\infty}^{1\textnormal{-SE}}$, we have $\sigma_2  ( \sigma_w \sqrt{ 8 {n} \ln  ( t  )   }  ) = o(t)$, and so it clearly follows that $\sigma_2  ( \overline{w}  (t,\delta  )  ) = o(t)$. This subsequently implies that $t \sigma_2  ( \overline{w}  (t,\delta  )  ) = o(t^2)$, and hence $ \ln ( \chi_4  ( t \sigma_2  ( \overline{w}(t,\delta)  )  )  ) = o(t)$ since $\chi_4 \in \mathcal{K}_{\infty}^{
    \textnormal{2\textnormal{-SE}}}$. Next, note that $\ln (\chi_3(t {(} \sigma_1(u_{\textnormal{max}}) {+1)} )) = o(t)$ holds, via $\chi_3 \in \mathcal{K}_{\infty}^{1\textnormal{-SE}}$ and
    {\versionedtext{Lem.~\ref{sysid:lemma:sub-exp-constant}}{\cite[Lem.~13]{siriya2024non}}.}
    Since we also have $\ln  ( \chi_1(t)  ) = o(t)$, using
    {\versionedtext{Lem.~\ref{sysid:lemma:sub-exp-sum}}{\cite[Lem.~12]{siriya2024non}}}
    we find that
    $
        \ln  ( \tilde{a}_1  ( \chi_1(t) + \chi_2 ( \vert x_0  \vert  ) + \chi_3  ( t \sigma_1  ( u_{\textnormal{max}}  ) ) +
          \chi_4  ( t \sigma_2  ( \overline{w}(t,\delta)  )  ) + c_1 + u_{\textnormal{max}}  )^{\tilde{a}_2}  )
        {\leq} \ln ( \tilde{a}_1  ) + \tilde{a}_2 \ln   ( \chi_1(t) + \chi_2 ( \vert x_0  \vert  ) + \chi_3  ( t {(} \sigma_1  ( u_{\textnormal{max}}  ) {+1)}  ) + \chi_4  ( t \sigma_2  ( \overline{w}(t,\delta)  )  ) + c_1 + u_{\textnormal{max}}  )
        = O(1) + o(t) = o(t).
    $
    {Clearly, $\ln(\tilde{a}_3+\sqrt{\gamma})=o(t)$. It then follows from
    {\versionedtext{Lem.~\ref{sysid:lemma:sub-exp-sum}}{\cite[Lem.~12]{siriya2024non}}}
    that $\ln  (\tilde{a}_1  (  \chi_1(t) + \chi_2 ( \vert x_0  \vert  ) + \chi_3  ( t ( \sigma_1  ( u_{\textnormal{max}}  ) + 1 )  ) + \chi_4  ( t \sigma_2  ( \overline{w}(t,\delta)  )  ) + c_1 + u_{\textnormal{max}}  )^{\tilde{a}_2} + \tilde{a}_3 + \sqrt{\gamma}  ) = o(t)$.}
    {Combined with \eqref{sysid:eqn:lemma-sub-exponential-gramian-1}, and the fact that $\liminf_{t \rightarrow \infty} \ln( \beta_{\textnormal{max}}(t,\delta,x_0))/t \geq \lim_{t \rightarrow \infty}\ln(\gamma)/t = 0$, it follows that $\beta_{\textnormal{max}}(\cdot,\delta,x_0)$ is $1$-SE.}
    {This concludes the proof.} \qed
\end{pf*}

{{Next}, we provide Lem.~\ref{sysid:lemma:data-dependent-error-bound}, which is a high-probability upper bound on the estimation error $\vert \hat{\theta}(t) - \theta_* \vert$ as a function of $\lambda_{\textnormal{min}}(G(t))$ and $\lambda_{\textnormal{max}}(G(t))$. It is similar to {the linear system results in} \cite[Thm.~1]{abbasi2011regret} and \cite[Lem.~4.1]{lale2022reinforcement},
{{but tailored to our linearly parametrised setting, and therefore} we defer the formal proof to \versionedtext{Sec.~\ref{sysid:sec:proof-data-dependent-error-bound}}{the Supplementary Results in \cite{siriya2024non}}.}}
\begin{lemma} \label{sysid:lemma:data-dependent-error-bound}
    (Data-Dependent Least-Squares Error Bounds)
    Suppose Ass.~\ref{sysid:assump:process-noise} is satisfied. Then, for any $\delta \in (0,1)$ and $x_0 \in \mathbb{X}$,
    $
        \mathbb{P}   ( \vert \hat{\theta}(t) - \theta_*  \vert \leq \frac{1}{ \lambda_{\textnormal{min}}  ( G(t)  )^{1/2} }  ( \gamma^{1/2}  {\vert \theta_* \vert} +
         \sigma_w  \sqrt{ {8}   ( \ln ({1} / \delta) + {n \ln (5)} + (d/2) \ln ( \lambda_{\textnormal{max}}(G(t)) \gamma^{-1} )  ) }  ), \ \forall t \in \mathbb{N}   )
        \geq 1 - \delta.
    $
\end{lemma}

{We now provide the proof of Thm.~\ref{sysid:theorem:rls-error-exciting}.}
\begin{pf*}{Proof of Thm.~\ref{sysid:theorem:rls-error-exciting}}
    Throughout this proof, suppose {$({\delta/3},x_0) \in \mathcal{C}$}.

    \sloppy
    Define the events ${\mathcal{E}_2},\mathcal{E}_3,\mathcal{E}_4 \in \mathcal{F}$ as {$\mathcal{E}_2:=\{ \lambda_{\textnormal{min}}  ( G(t) ) \geq q(t-1) + \gamma, \forall t \in \mathcal{T}_{\textnormal{PE}}({\delta/3},x_0) \}$}, $\mathcal{E}_3 :=  $\\$\big \{  \vert \hat{\theta}(t) - \theta_*  \vert \leq \frac{1}{ \lambda_{\textnormal{min}}  ( G(t)  )^{1/2} } $ $( \sigma_w \sqrt{ {8}   ( \ln (3 / \delta) + {n \ln (5)} + (d/2) \ln ( \lambda_{\textnormal{max}}(G(t)) \gamma^{-1} )  ) } + \gamma^{1/2}  \vert \theta_*  \vert_F ), \ \forall t \in \mathbb{N}  \big \}$ and $\mathcal{E}_4 :=  \{  \vert X(t-1)  \vert \leq \overline{x}(t-1,\delta/3,x_0), \  \vert Z(t)  \vert \leq  \overline{z}(t,\delta/3,x_0) , \ \forall t \in \mathbb{N}  \}$ respectively.
    Next, notice that on the event $\mathcal{E}_4$, for all $t \in \mathbb{N}$, $\lambda_{\textnormal{max}}  ( G(t) ) = \lambda_{\textnormal{max}}  ( \sum_{i=1}^t Z(i) Z(i)^{\top} + \gamma I ) \leq \lambda_{\textnormal{max}} ( \sum_{i=1}^t Z(i) Z(i)^{\top}  ) + \gamma \leq \sum_{i=1}^t  \vert Z(i)  \vert^2 + \gamma \leq \sum_{i=1}^t \overline{z}^2(i,\delta/3,x_0) + \gamma = \beta_{\textnormal{max}}(t,\delta/3,x_0)$.
    {It follows that} on the event $\mathcal{E}_2 \cap \mathcal{E}_3 \cap \mathcal{E}_4$, $ \vert \hat{\theta}(t) - \theta_*  \vert \leq e(t,\delta,x_0), \ \forall t \in \mathcal{T}_{\textnormal{PE}}({\delta/3},x_0)$,
    with $e$ from \eqref{sysid:eqn:theorem-rls-error-exciting-e}. Thus, we can establish that
    \begin{align}
        &\mathbb{P} \mleft ( \mleft \vert \hat{\theta}(t) - \theta_* \mright \vert \leq e(t,{\delta},x_0),\ \forall t \in {\mathcal{T}_{\textnormal{PE}}({\delta/3},x_0)} \mright ) \\
        &\geq \mathbb{P} \mleft ( \mathcal{E}_3 \cap \mathcal{E}_4 \cap \mathcal{E}_2 \mright ) \geq 1 - {\delta}  \label{sysid:eqn:theorem-rls-error-exciting-5}
    \end{align}
    The second inequality in \eqref{sysid:eqn:theorem-rls-error-exciting-5} follows by applying the union bound, and the fact that $\mathbb{P}(\mathcal{E}_3) \geq 1 - \delta/3$ from Lem.~\ref{sysid:lemma:data-dependent-error-bound}, $\mathbb{P}(\mathcal{E}_4) \geq 1 - \delta/3$ from Lem.~\ref{sysid:lemma:high-prob-state-bound}, and $\mathbb{P}(\mathcal{E}_2) \geq 1 - {\delta/3}$ from {Ass.~\ref{assump:interval-pe}}.

    {Since $\beta_{\textnormal{max}}(t,\delta/3,x_0)$ is $1$-SE from Prop.~\ref{prop:sub-exp-gramian},} this implies {$\lim_{t \rightarrow \infty} \frac{ \ln ( \beta_{\textnormal{max}}(t,{\delta/3},x_0) )}{ q(t-1)+\gamma} = 0$}, and so it follows from the definition of $e$ in \eqref{sysid:eqn:theorem-rls-error-exciting-e} that
    $
        \lim_{t \rightarrow \infty} e(t,\delta,x_0) = 0.
    $
\qed \end{pf*}

\subsubsection{Proofs of Supporting Results} \label{sysid:sec:proof-supporting-results}

{The proof of Lem.~\ref{sysid:lemma:high-prob-state-bound} is} provided here.

\begin{pf*}{Proof of Lem.~\ref{sysid:lemma:high-prob-state-bound}}
    Firstly, note that for all $t \in {\mathbb{N}_0}$, $X(t) = \phi \mleft (t,x_0, \{ U(i) \}_{i = 0}^{t-1}, \{ W(i) \}_{i=1}^t \mright )$.
    Suppose $\delta \in (0,1)$. Let
    \begin{align}
        \tilde{\mathcal{E}} :=  \mleft \{ \mleft \vert W(t) \mright \vert \leq \overline{w}(t,\delta), \ \forall t \in \mathbb{N} \mright \}, \label{sysid:eqn:lemma-high-prob-state-bound-1}
    \end{align}
    \sloppy
    which satisfies $\mathbb{P} ( \tilde{\mathcal{E}}  ) \geq 1 - \delta$ by {\versionedtext{Lem.~\ref{sysid:lemma:sub-gaussian-uniform-bound}}{\cite[Lem.~7]{siriya2024non}}}
    and Ass.~\ref{sysid:assump:process-noise}.
    Due to Ass.~\ref{sysid:assump:constrained-controls} and \ref{sysid:assump:forward-complete}, and \eqref{sysid:eqn:lemma-high-prob-state-bound-1}, on the event $\tilde{\mathcal{E}}$,
    $
        {\lvert} X(t) {\rvert} = {\lvert} \phi \big (t,x_0, \{ U(i) \}_{i = 0}^{t-1}, \{ W(i) \}_{i=1}^t \big ) {\rvert}
        \leq \chi_1(t) + \chi_2( \vert x_0  \vert ) + \chi_3 \Big( \sum_{i=0}^{t-1}  \sigma_1  (  \vert U(i)  \vert  ) \Big )
        + \chi_4 \Big( \sum_{i=1}^t  \sigma_2  (  \vert W(i)  \vert  ) \Big ) + c_1  \leq
        \overline{x}(t,\delta,x_0)
    $
    for all $t \in {\mathbb{N}_0}$. Then, using Ass.~\ref{sysid:assump:poly-feature-map}, on the event $\tilde{\mathcal{E}}$, we have
    $
        \vert Z(t) \vert = \vert \psi (X(t-1),U(t-1)) \vert
        \leq \chi_5 \big (  \big \vert [X(t-1)^{\top} \ U(t-1)^{\top}]^{\top} \big \vert  \big ) \leq
        \overline{z}(t,\delta,x_0)
    $
    for all $t \in \mathbb{N}$.
\qed \end{pf*}

\subsection{{Proofs for Certifying Relaxed PE}} \label{sec:proof-certify-excitation}

{In this section, we provide {Lem.~\ref{sysid:lemma:regional-pe}, which} Prop.~\ref{prop:regional-excitation-pe} directly relies on, and the proof of Prop.~\ref{prop:regional-excitation-pe} and Cor.~\ref{sysid:corollary:rls-error-global-poly}.
{The proof of Cor.~\ref{sysid:corollary:rls-error-global} is omitted as it {immediately} follows from Thm.~\ref{sysid:theorem:rls-error-exciting} and Prop.~\ref{prop:regional-excitation-pe}}.
The proof of Lem.~\ref{sysid:lemma:regional-pe} is deferred to Sec.~\ref{sysid:sec:proof-supporting-results}.}

\sloppy
{{Firstly}, we need to probabilistically lower bound $\lambda_{\textnormal{min}}(\sum_{i=1}^t Z(i) Z(i)^{\top})$. This is enabled via the use of Lem.~\ref{sysid:lemma:regional-pe}. It says that there exists a high-probability event $\mathcal{E}_2$, such that for all time steps $t \geq T_{\textnormal{burn-in}}(3\delta,x_0)$, if 1) our sample belongs to this event, 2) the upper bound $ \vert Z(i)  \vert \leq  \overline{z}(i,\delta,x_0)$ holds on the regressor for all $i \leq t$, and 3) the one-step predicted state $\mathbb{E}[X(i) \mid X(i-1), U(i-1) ]$ remains in $\mathcal{X}_{\textnormal{PE}}$ for all $i \leq t-1$, then a linearly increasing lower bound on $\lambda_{\textnormal{min}} (\sum_{i = 1}^t Z(i) Z(i)^{\top} )$ holds --- i.e., PE is holding. We interpret this as a \textit{regional PE} result due to the requirement that the one-step predicted state remains in $\mathcal{X}_{\textnormal{PE}}$. Note that conditioning on the event $\mathcal{E}_2$ is required due to the stochastic nature of the problem, in the sense that PE also depends on the injected and process noise affecting the system.
}

\begin{lemma} \label{sysid:lemma:regional-pe}
    (Regional Persistency of Excitation)
    Suppose Ass.~\ref{sysid:assump:process-noise}, \ref{sysid:assump:forward-complete}, \ref{sysid:assump:constrained-controls}, \ref{sysid:assump:poly-feature-map}
    {and \ref{sysid:assump:regional-excitation} are satisfied.}
    Moreover, suppose that $(\psi,\alpha,\mu_s,\mu_w)$ is {$(\mathcal{X}_{\textnormal{PE}},c_{\textnormal{PE}},p_{\textnormal{PE}})$} for some $\mathcal{X}_{\textnormal{PE}} \subseteq \mathbb{X}$, {$c_{\textnormal{PE}},p_{\textnormal{PE}} > 0$}.
    Then, for any $x_0 \in \mathbb{X}$ and $\delta \in (0,{1/2})$, there exists an event $\mathcal{E}_2 \in \mathcal{F}$ satisfying $\mathbb{P}(\mathcal{E}_2) \geq 1 - \delta$, such that for any $t \geq T_{\textnormal{burn-in}}({2}\delta,x_0)$,
   \begin{align}
        &\lambda_{\textnormal{min}} \mleft(\sum_{i = 1}^t Z(i) Z(i)^{\top} \mright) \geq \frac{{{c_{\textnormal{PE}}} p_{\textnormal{PE}}}}{4}(t-1) \text{ on the event}\\
        & \mathcal{E}_2 \cap \mleft \{ \mleft \vert Z(i) \mright \vert \leq  \overline{z}\mleft(i,\delta,x_0\mright), \ \forall i \leq t \mright \} \cap \\
        & \ \{ \mathbb{E}[X(i) \mid X(i-1), U(i-1) ] \in \mathcal{X}_{\textnormal{PE}}, \ \forall i \leq t-1  \}. \quad \label{sysid:eqn:lemma-regional-pe-main}
    \end{align}
\end{lemma}

{Prop.~\ref{prop:regional-excitation-pe} can now be established.}

\begin{pf*}{Proof of Prop.~\ref{prop:regional-excitation-pe}}
    {
        Let $\mathcal{C} \subseteq (0,1) \times \mathbb{X} $, $q > 0$, and $\mathcal{T}_{\textnormal{PE}}: {\mathcal{C}} \rightarrow \mathbb{I}$, be as stated in Prop.~\ref{prop:regional-excitation-pe}.
        Suppose $(\delta,x_0) \in \mathcal{C}$.
    }
    Define the event $\mathcal{E}_4 :=  \{  \vert X(t-1)  \vert \leq \overline{x}(t-1,\delta/{2},x_0), \  \vert Z(t)  \vert \leq  \overline{z}(t,\delta/{2},x_0) , \ \forall t \in \mathbb{N}  \}$.
    Moreover, let $\mathcal{E}_2 \in \mathcal{F}$ be an event such that for all $t \geq T_{\textnormal{burn-in}}({\delta},x_0)$,
    \begin{align}
        &\lambda_{\textnormal{min}} \mleft(\sum_{i = 1}^t Z(i) Z(i)^{\top} \mright) \geq \frac{{{c_{\textnormal{PE}}} p_{\textnormal{PE}}}}{4}(t-1) \ \text{on the event} \\
        & \ \mathcal{E}_2 \cap \mleft \{ \mleft \vert Z(i) \mright \vert \leq  \overline{z}\mleft(i,\delta/{2},x_0\mright), \ \forall i \leq t \mright \} \cap \\
        & \ \{ \mathbb{E}[X(i) \mid X(i-1), U(i-1) ] \in \mathcal{X}_{\textnormal{PE}}, \ \forall i \leq t-1  \}, \label{sysid:eqn:theorem-rls-error-exciting-6}
    \end{align}
    and $\mathbb{P}(\mathcal{E}_2) \geq 1 - \delta/{2}$, where the existence of a satisfactory $\mathcal{E}_2$ is established in Lem.~\ref{sysid:lemma:regional-pe}.
    Notice that on the event $\mathcal{E}_4$, for all $t \in \mathbb{N}$, $\mathbb{E} \mleft [ X(t) \mid X(t-1), U(t-1) \mright ] = g(X(t-1),U(t-1),0) \in \Gamma \mleft ( B_{\overline{x}(t-1,\delta/{2},x_0)}(0) {\cap \mathbb{X}} \mright ) $, which {alongside monotonicity of $\Gamma$ and $\bar{x}$} implies that for all $t \leq T_{\textnormal{excited}}({\delta},x_0) - 1$, $\mathbb{E} [ X(t) \mid X(t-1), U(t-1) ]  \in \mathcal{X}_{\textnormal{PE}}$.
    Combining this fact with \eqref{sysid:eqn:theorem-rls-error-exciting-6}, it follows that on the event $\mathcal{E}_4 \cap \mathcal{E}_2$, for all {$t \in \mathcal{T}_{\textnormal{PE}}(\delta,x_0)$},
    \begin{align}
        &\lambda_{\textnormal{min}} \mleft ( G(t) \mright) = \lambda_{\textnormal{min}} \mleft ( \sum_{i=1}^t Z(i) Z(i)^{\top} + \gamma I \mright) \\
        &\geq \frac{{c_{\textnormal{PE}}}p_{\textnormal{PE}}}{4}(t-1) + \gamma . \label{sysid:eqn:theorem-rls-error-exciting-3}
    \end{align}
    {Satisfaction of Ass.~\ref{assump:interval-pe}} follows by applying the union bound, and from the fact that $\mathbb{P}(\mathcal{E}_4) \geq 1 - \delta/{2}$ from Lem.~\ref{sysid:lemma:high-prob-state-bound}, and $\mathbb{P}(\mathcal{E}_2) \geq 1 - \delta/{2}$ from earlier in this proof. {Thm.~\ref{sysid:theorem:rls-error-exciting} then also holds since Ass.~\ref{assump:interval-pe} holds.}

    We now establish that $T_{\textnormal{burn-in}}({\delta},x_0) < \infty$.
    {Note that $\ln \Big (1 + \frac{16 \sum_{i=1}^t \overline{z}^2  (i, \delta/{2}, x_0  ) }{ {c_{\textnormal{PE}}} p_{\textnormal{PE}} (t-1) } \Big) \leq\ln(1+ \tfrac{16}{c_{\textnormal{PE}}p_{\textnormal{PE}}} \beta_{\textnormal{max}}(t,\delta/2,x_0)) = o(t)$ due to Prop.~\ref{prop:sub-exp-gramian} and
    \versionedtext{Lem.~\ref{sysid:lemma:sub-exp-sum}}{\cite[Lem.~12]{siriya2024non}},}
    and so it follows that
    $
        \frac{2}{(1 - \ln(2))  p_{\textnormal{PE}}} \Big( \ln  ( \frac{\pi^2 t^2 }{{3}\delta}  )
          + d \ln  \Big( 1 + \frac{16 {\sum_{i=1}^t \overline{z}^2 (i, \delta/{2}, x_0 )} }{ c_{\textnormal{PE}} p_{\textnormal{PE}} (t-1) }  \Big)   \Big) + 1 = o(t).
    $
    Combined with the fact that $(t-T+1)^2\leq t^2$ for all $T\in \mathbb{N}$ and $t \geq \max(2,{T})$,
    {it immediately follows that there exists an integer $\overline{T} \geq 2$ such that for all $t \geq \overline{T}$,
    $
        t \geq \frac{2}{(1 - \ln(2))  p_{\textnormal{PE}}} \Big( \ln   ( \frac{\pi^2 (t - \overline{T} + 1)^2 }{{3}\delta}   )  + d \ln \Big(1 + \frac{16 \sum_{i=1}^t \overline{z}^2  (i, \delta/{2}, x_0 ) }{ c_{\textnormal{PE}} p_{\textnormal{PE}} (t-1) }  \Big)  \Big) + 1
    $
    and hence $T_{\textnormal{burn-in}}(\delta,x_0) \leq \overline{T} < \infty$.}
\qed \end{pf*}

\begin{pf*}{Proof of Cor.~\ref{sysid:corollary:rls-error-global-poly}}
    Suppose $x_0 \in \mathbb{X}$ and $\delta \in (0,1)$. Since Ass.~\ref{sysid:assump:poly-forward-complete} implies Ass.~\ref{sysid:assump:forward-complete},
    the results in Cor.~\ref{sysid:corollary:rls-error-global} automatically hold. We are left to prove that $e(t,{\delta},x_0) = O ( \sqrt{ \ln(t)/t }  )$.

    Following the same steps as {Prop.~\ref{prop:sub-exp-gramian}}, similarly to \eqref{sysid:eqn:lemma-sub-exponential-gramian-3} we know there exist $\tilde{a}_1, \tilde{a}_2, \tilde{a}_3 \geq 0 $ such that for all $t \in \mathbb{N}$,
    $
        \overline{z}^2  (t, \delta/3, x_0  ) \leq \tilde{a}_1  (  \chi_1(t) + \chi_2 ( \vert x_0  \vert  ) + \chi_3  ( t \sigma_1  ( u_{\textnormal{max}}  )  )
        + \chi_4  ( t \sigma_2  ( \overline{w}(t,\delta/3 )  )  ) + c_1 + u_{\textnormal{max}}  )^{\tilde{a}_2} + \tilde{a}_3.
    $
    Recall from Ass.~\ref{sysid:assump:poly-forward-complete} that {$\chi_1,\chi_3,\chi_4,\sigma_2$ are APB functions}. Since $\overline{w}(t,\delta/3) = O(t)$, it follows that $\sigma_2 (\overline{w}(t,\delta/3))$ {is APB in $t$}, which implies $\chi_4 \mleft ( t \sigma_2 \mleft ( \overline{w}\mleft(t,\delta/3 \mright) \mright ) \mright )$ and $  \chi_1(t) + \chi_2\mleft (\mleft \vert x_0 \mright \vert \mright ) + \chi_3 \mleft ( t \sigma_1 \mleft ( u_{\textnormal{max}} \mright ) \mright ) + \chi_4 \mleft ( t \sigma_2 \mleft ( \overline{w}\mleft(t,\delta/3 \mright) \mright ) \mright ) + c_1 + u_{\textnormal{max}}$ {is as well}. This implies that $\beta_{\textnormal{max}}({t,}\delta/3,x_0) = {\sum_{i=1}^t} \overline{z}^2 \mleft ({i}, \delta/3, x_0 \mright ) + \gamma {\leq} {t} \overline{z}^2 \mleft ({t}, \delta/3, x_0 \mright ) + \gamma$ {is APB}, and so $\ln(\beta_{\textnormal{max}}(t,\delta/3,x_0){)} = O \mleft ( \ln (t) \mright )$.
    {It then follows that}
    $
        e(t,{\delta},x_0) = O ( \sqrt{ \ln(t)/t }  ) .
    $
\qed \end{pf*}

\subsubsection{Proofs for Supporting Results}

{{The main focus of this section is} proving Lem.~\ref{sysid:lemma:regional-pe}. Before doing so, we provide an intermediate result in Lem. \ref{sysid:lemma:single-direction-regional-pe}. It says that there exists a high-probability event $\mathcal{E}_1$, such that if 1) our sample belongs to this event, and 2) the one-step predicted state $\mathbb{E}[X(i) \mid X(i-1), U(i-1) ]$ remains in $\mathcal{X}_{\textnormal{PE}}$ for all $i \leq t-1$, then for all directions $v \in \mathcal{S}^{d-1}$ inside the unit sphere, a linearly increasing lower bound on $\sum_{i = 1}^t  \vert v^{\top} Z(i)  \vert^2$ holds. This result can be interpreted as a \textit{single-direction regional PE} result, since we obtain a linearly increasing bound on the projection of $\sum_{i=1}^t Z(i)Z(i)^{\top}$ onto each direction of the unit sphere.
{Here,} $\mathcal{E}_1$ is required due to the stochastic nature of the problem. {This result} is analogous to \cite[Prop.~2.5]{simchowitz2018learning}, the key difference being that we provide a lower bound only when $\mathbb{E}[X(i) \mid X(i-1), U(i-1) ]$ remains in $\mathcal{X}_{\textnormal{PE}}$ due to the regional excitation assumption.}

\begin{lemma} \label{sysid:lemma:single-direction-regional-pe}
    (Single-Direction Regional PE)
    Suppose Ass.~\ref{sysid:assump:process-noise}
    {and \ref{sysid:assump:regional-excitation} are satisfied.}
    Then, for any $x_0 \in \mathbb{X}$, $v \in \mathcal{S}^{d-1}$ and $t \in \mathbb{N}$, there exists an event $\mathcal{E}_1 \in \mathcal{F}$ such that $\mathbb{P}(\mathcal{E}_1) \geq 1 - e^{-\frac{1}{2}({1-\ln(2)}){p_{\textnormal{PE}}}(t-1)}$ and
    \begin{align}
         &\sum_{i = 1}^t \mleft \vert v^{\top} Z(i) \mright \vert^2 \geq \frac{{{c_{\textnormal{PE}}} p_{\textnormal{PE}}}}{2} (t-1) \quad \text{on the event} \quad \mathcal{E}_1 \cap
         \\
         & \quad \{ \mathbb{E}[X(i) \mid X(i-1), U(i-1) ] \in \mathcal{X}_{\textnormal{PE}}, \\
         & \qquad \forall i \in \{ 1, \hdots, t-1 \}  \}. \label{sysid:eqn:lemma-single-direction-regional-pe-main}
    \end{align}
\end{lemma}
\begin{pf}
    Suppose $v \in \mathcal{S}^{d-1}$, $x_0 \in \mathbb{X}$ and $t \in \mathbb{N}$. Denote $\tilde{\mathcal{E}}({i}) =  \{ \mathbb{E}[X(j) \mid X(j-1), U(j-1) ] \in \mathcal{X}_{\textnormal{PE}}, \ \forall j \in \{ 1, \hdots, i \}  \}, \ i \in \mathbb{N}_{{0}}$
    and let
    $\mathcal{E}_1 = \tilde{\mathcal{E}}^{\comp}(t-1) \cup  \{ \sum_{i = 1}^t \vert v^{\top} Z(i) \vert^2 \geq \frac{{{c_{\textnormal{PE}}} p_{\textnormal{PE}}}}{2} (t-1)  \}$.

    We first verify that \eqref{sysid:eqn:lemma-single-direction-regional-pe-main} holds:
    $
        \mathcal{E}_1 \cap \tilde{\mathcal{E}}(t-1)
        = \tilde{\mathcal{E}}(t-1) \cap  \{ \sum_{i = 1}^t \vert v^{\top} Z(i) \vert^2 \geq \frac{{{c_{\textnormal{PE}}} p_{\textnormal{PE}}}}{2} (t-1)  \}
        \subseteq  \{ \sum_{i = 1}^t \vert v^{\top} Z(i) \vert^2 \geq \frac{{{c_{\textnormal{PE}}} p_{\textnormal{PE}}}}{2} (t-1)  \}.
    $

    We now turn our attention to verifying that $\mathbb{P}(\mathcal{E}_1) \geq 1 - e^{-\frac{1}{2}({1-\ln(2)}){p_{\textnormal{PE}}}(t-1)}$.
    Note that $\sum_{i=1}^t  \vert v^{\top} Z(i)  \vert^2 \geq {c_{\textnormal{PE}}} \sum_{i=1}^t \mathbf{1}_{ \{  \vert v^{\top} Z(i)  \vert^2 \geq {c_{\textnormal{PE}}}  \}}$ holds trivially, and therefore
    \begin{align}
        &\mleft \{ \sum_{i=1}^t \mleft \vert v^{\top} Z(i) \mright \vert^2 \leq \frac{{c_{\textnormal{PE}}} p_{\textnormal{PE}}}{2}(t-1) \mright \} \\
        &\subseteq \mleft \{ \sum_{i=1}^t \mathbf{1}_{\mleft \{ \mleft \vert v^{\top} Z(i) \mright \vert^2 \geq {c_{\textnormal{PE}}} \mright \}} \leq \frac{p_{\textnormal{PE}}}{2}(t-1) \mright \}. \label{sysid:eqn:lemma-scalar-min-eigenvalue-bound-2}
    \end{align}
    Next, note that the following result holds:
    \begin{align}
        & \mathbb{P} \mleft ( \mathcal{E}_1^{\comp} \mright )
        \leq \mathbb{P} \Bigg( \tilde{\mathcal{E}}(t-1) \cap \\
        & \quad \mleft. \mleft \{ \sum_{i = 1}^t \mathbf{1}_{\mleft \{ \vert v^{\top} Z(i) \vert^2 \geq {c_{\textnormal{PE}}} \mright \}} \leq \frac{p_{\textnormal{PE}}}{2} (t-1) \mright \}  \mright) \label{sysid:eqn:lemma-scalar-min-eigenvalue-bound-4} \\
        &\leq \inf_{s<0} \mathbb{E}\mleft [ \mathbf{1}_{\tilde{\mathcal{E}}(t-1)} e^{s \sum_{i=1}^t \mathbf{1}_{\mleft \{ \vert v^{\top} Z(i) \vert^2 \geq {c_{\textnormal{PE}}} \mright \}}}  \mright ] e^{-s \frac{p_{\textnormal{PE}}}{2} (t-1)} \label{sysid:eqn:lemma-scalar-min-eigenvalue-bound-5}
    \end{align}
    where \eqref{sysid:eqn:lemma-scalar-min-eigenvalue-bound-4} follows from \eqref{sysid:eqn:lemma-scalar-min-eigenvalue-bound-2}, and \eqref{sysid:eqn:lemma-scalar-min-eigenvalue-bound-5} follows from the joint Chernoff bound in {\versionedtext{Lem.~\ref{sysid:lemma:joint-chernoff-bound}}{\cite[Lem.~8]{siriya2024non}}}.

    Let $\mathcal{G}(i) = \sigma(\{X(j)\}_{j \leq i-1}, \{U(j)\}_{j \leq i-1})$ for $i \in \mathbb{N}$ with $\mathcal{G}(0)$ the trivial sigma field. Next, we upper bound the expectation in \eqref{sysid:eqn:lemma-scalar-min-eigenvalue-bound-5}. For all $s < 0$ and $i \in \mathbb{N}$, we have
    \begin{align}
        & \mathbb{E} \mleft[ \mathbf{1}_{\tilde{\mathcal{E}}(i-1)} e^{s \sum_{j=1}^i \mathbf{1}_{\mleft \{ \vert v^{\top} Z(j) \vert^2 \geq c_{\textnormal{PE}} \mright \}}  }  \mright]\\
        &= \mathbb{E}\mleft [  \mathbf{1}_{\tilde{\mathcal{E}}(i-1)} e^{s \sum_{j=1}^{i-1} \mathbf{1}_{\mleft \{ \mleft \vert v^{\top} Z(j) \mright \vert^2 \geq c_{\textnormal{PE}} \mright \}}  }  \mright. \\
        & \qquad \mleft. \times \mathbb{E}\mleft [ e^{s \mathbf{1}_{\mleft \{ \mleft \vert v^{\top} Z(i) \mright \vert^2 \geq c_{\textnormal{PE}} \mright \}} } \mid \mathcal{G}(i-1) \mright ] \mright ], \label{sysid:eqn:lemma-scalar-min-eigenvalue-bound-6}
    \end{align}
    where \eqref{sysid:eqn:lemma-scalar-min-eigenvalue-bound-6} follows from the tower property of conditional expectation and since for all $j \leq i - 1$, $Z(j)$ is $\mathcal{G}(i-1)$-measurable. Then, using the law of total expectation,
    $
        \mathbb{E}  [ e^{s \mathbf{1}_{ \{  \vert v^{\top} Z(i)  \vert^2 \geq {c_{\textnormal{PE}}}  \}} } \mid \mathcal{G}(i-1)  ]
        = 1 - (1 - e^{s}) \mathbb{P}  (  \vert v^{\top} Z(i)  \vert^2 \geq {c_{\textnormal{PE}}} \mid \mathcal{G}(i-1)  ),
    $
    for all $s < 0$ and $i \in \mathbb{N}$.
    Combining with \eqref{sysid:eqn:lemma-scalar-min-eigenvalue-bound-6}, for all $t \in \mathbb{N}$ and $s < 0$,
    \begin{align}
        &\mathbb{E} \mleft[ \mathbf{1}_{\tilde{\mathcal{E}}({t}-1)} e^{s \sum_{j=1}^{{t}} \mathbf{1}_{\mleft \{ \vert v^{\top} Z(j) \vert^2 \geq c_{\textnormal{PE}} \mright \}}  }  \mright] \\
        &= \mathbb{E} \mleft [  \mathbf{1}_{\tilde{\mathcal{E}}(t-1)} e^{s  \sum_{i=1}^{t-1} \mathbf{1}_{\mleft \{ \mleft \vert v^{\top} Z(i) \mright \vert^2 \geq c_{\textnormal{PE}} \mright \}}  } \mright. \\
        & \quad \mleft. \times \mleft ( 1 - \mleft (1 - e^s \mright ) P \mleft ( \mleft \vert v^{\top} Z(t) \mright \vert^2 \geq c_{\textnormal{PE}} \mid \mathcal{G}(t-1) \mright ) \mright ) \mright ] \label{sysid:eqn:lemma-scalar-min-eigenvalue-bound-14} \\
        &\leq \mathbb{E} \mleft [  e^{s \mathbf{1}_{ \{  \vert v^{\top} Z(1)  \vert^2 \geq c_{\textnormal{PE}}  \}}  } \mright ] \mleft ( 1 - \mleft (1 - e^s \mright ) p_{\textnormal{PE}} \mright )^{t-1} \label{sysid:eqn:lemma-scalar-min-eigenvalue-bound-17} \\
        &\leq \mleft ( 1 - \mleft (1 - e^s \mright ) p_{\textnormal{PE}} \mright )^{t-1} \label{sysid:eqn:lemma-scalar-min-eigenvalue-bound-16}
    \end{align}
    {where} \eqref{sysid:eqn:lemma-scalar-min-eigenvalue-bound-17} follows {from Ass.~\ref{sysid:assump:regional-excitation}, $\mathbf{1}_{\tilde{\mathcal{E}}(t)} \leq \mathbf{1}_{\tilde{\mathcal{E}}(t-1)}$ for all $t \in \mathbb{N}$, and} $t-1$ iterations, and {\eqref{sysid:eqn:lemma-scalar-min-eigenvalue-bound-16} holds} since $\mathbb{E}[  e^{s \mathbf{1}_{\{ \vert v^{\top} Z(1) \vert^2 \geq {c_{\textnormal{PE}}} \}}  } ] \leq 1$.

    Finally, by combining \eqref{sysid:eqn:lemma-scalar-min-eigenvalue-bound-5} and \eqref{sysid:eqn:lemma-scalar-min-eigenvalue-bound-16}, and using
    {\versionedtext{Lem.~\ref{sysid:lemma:simplify-min-eigenvalue-bound}}{\cite[Lem.~10]{siriya2024non}}},
    we have
    $
        \mathbb{P}  ( \tilde{\mathcal{E}}(t-1)\cap  \{ \sum_{i = 1}^t \vert v^{\top} Z(i) \vert^2 \leq \frac{{c_{\textnormal{PE}}} p_{\textnormal{PE}}}{2} (t-1)  \}  )
        \leq \inf_{s<0}  (p_{\textnormal{PE}} e^{s(1-\frac{p_{\textnormal{PE}}}{2})} + \mleft ( 1-p_{\textnormal{PE}} \mright )e^{-s \frac{p_{\textnormal{PE}}}{2}}  )^{t-1}
        \leq e^{-\frac{1}{2} \mleft ( 1 - \ln(2) \mright )p_{\textnormal{PE}}(t-1)}
    $
    for all $t \in \mathbb{N}$.
\qed \end{pf}
We now move from the single-direction regional PE result in Lem.~\ref{sysid:lemma:single-direction-regional-pe} to Lem.~\ref{sysid:lemma:regional-pe}, which lower bounds $\sum_{i=1}^t Z(i) Z(i)^{\top}$ uniformly in all directions, using a \textit{covering argument.}
\begin{remark}
    {Covering arguments are similarly used to establish PE in \cite{simchowitz2018learning} (see proof of Thm.~2.4). We refer to \cite{matni2019tutorial} for a tutorial showing how this {is} done. {The main difference is} we derive a result that says PE holds when $g(X(t-1),U(t-1),0)$ remains inside an \textit{excited region}.}
\end{remark}

\begin{pf*}{Proof of Lem.~\ref{sysid:lemma:regional-pe}}
    Suppose $x_0 \in \mathbb{X}$, $\delta \in (0,{1/2})$.

    {If $\sum_{i=1}^t \overline{z}^2\mleft (i,\delta,x_0 \mright) > 0$}, let $\mathcal{C}(t)$ be a family of $\epsilon(t)$-covers
    {\cite[Def.~20.6]{lattimore2020bandit}}
    for $\mathcal{S}^{d-1}$ parametrised over {$t \geq 2$}, {where}
    \begin{align}
        \epsilon(t) := \frac{{c_{\textnormal{PE}}}p_{\textnormal{PE}}(t-1)}{8 \sum_{i=1}^t \overline{z}^2\mleft (i,\delta,x_0 \mright)}, \ t {\geq 2}, \label{sysid:eqn:lemma-regional-pe-1}
    \end{align}
    {and satisfying} $\mathrm{card}(\mathcal{C}(t){)} \leq (1 + 2/\epsilon(t))^d$, whose existence is verified in
    {\cite[Cor.~4.2.13]{vershynin2018high}}.
    {If $\sum_{i=1}^t \overline{z}^2\mleft (i,\delta,x_0 \mright) = 0$, let $\mathcal{C}(t):=\{v_0\}$ for some $v_0 \in \mathcal{S}^{d-1}$.}
    Let $\mathcal{E}_1(t,v)$ be a family of events in $\mathcal{F}$ satisfying \eqref{sysid:eqn:lemma-single-direction-regional-pe-main} and $\mathbb{P}(\mathcal{E}_1(t,v)) \geq 1 - e^{-\frac{1}{2}({1-\ln(2)}){p_{\textnormal{PE}}}(t-1)}$ for all $t \in \mathbb{N}$ and $v \in \mathcal{S}^{d-1}$, with existence established in Lem.~\ref{sysid:lemma:single-direction-regional-pe}. Let
    \begin{align}
        \tilde{\mathcal{E}}_1(t) := \mleft \{ \mleft \vert Z(i) \mright \vert \leq  \overline{z}\mleft (i,\delta,x_0 \mright) , \ \forall i \leq t \mright \}, \ t \in \mathbb{N}. \label{sysid:eqn:lemma-regional-pe-2}
    \end{align}
    Moreover, let $\mathcal{E}_2 := \bigcap_{t \geq T_{\textnormal{burn-in}}} \bigcap_{v \in \mathcal{C} \mleft (t \mright )} \mathcal{E}_1(t,v)$.
    We now show that $\mathcal{E}_2$ satisfies \eqref{sysid:eqn:lemma-regional-pe-main}. Denote $\tilde{\mathcal{E}}_3(t) := \{ \mathbb{E}[X(i) \mid X(i-1), U(i-1)]  \in \mathcal{X}_{\textnormal{PE}}, \ \forall i \leq t-1  \}$ for $t \in \mathbb{N}$.
    Then, for all $t \in \mathbb{N}$ and $v \in \mathcal{S}^{d-1}$, on the event $\tilde{\mathcal{E}}_3(t) \cap \mathcal{E}_1(t,v)$,
    \begin{align}
        \sum_{i = 1}^t \mleft \vert v^{\top} Z(i) \mright \vert^2 \geq \frac{{{c_{\textnormal{PE}}} p_{\textnormal{PE}}}}{2} (t-1) \label{sysid:eqn:lemma-regional-pe-3}
    \end{align}
    due to Lem.~\ref{sysid:lemma:single-direction-regional-pe}. It then follows that for all $t \geq T_{\textnormal{burn-in}}$ ({denoting} $T_{\textnormal{burn-in}} = T_{\textnormal{burn-in}}({{2}\delta},x_0)$ {for simplicity}), {if $\sum_{i=1}^t \overline{z}^2\mleft (i,\delta,x_0 \mright) > 0$,} on the event $\mathcal{E}_2 \cap \tilde{\mathcal{E}}_1( t ) \cap \tilde{\mathcal{E}}_3(t) $,
    \begin{align}
        &\lambda_{\text{min}} \mleft (\sum_{i=1}^t Z(i) Z(i)^{\top} \mright )
        = \inf_{ u \in \mathcal{S}^{d-1} } \sum_{i = 1}^t \mleft \vert u^{\top} Z(i) \mright \vert^2 \\
        &\geq \min_{v \in \mathcal{C}(t)} \sum_{i = 1}^t \mleft \vert v^{\top} Z(i) \mright \vert^2 - 2 \mleft \vert \sum_{i=1}^t Z(i) Z(i)^{\top} \mright \vert \epsilon(t) \label{sysid:eqn:lemma-regional-pe-4} \\
        &\geq
        \frac{{c_{\textnormal{PE}}}p_{\textnormal{PE}}}{4}(t-1) \label{sysid:eqn:lemma-regional-pe-5}
    \end{align}
    where \eqref{sysid:eqn:lemma-regional-pe-4} follows since $\mathcal{C}(t)$ is an $\epsilon(t)$-cover, and \eqref{sysid:eqn:lemma-regional-pe-5} follows from \eqref{sysid:eqn:lemma-regional-pe-1}, \eqref{sysid:eqn:lemma-regional-pe-2}, and \eqref{sysid:eqn:lemma-regional-pe-3}, thus verifying \eqref{sysid:eqn:lemma-regional-pe-main}.
    {On the other hand, if $\sum_{i=1}^t \overline{z}^2\mleft (i,\delta,x_0 \mright) = 0$, then \eqref{sysid:eqn:lemma-regional-pe-5} holds vacuously since $\mathcal{E}_2\cap\tilde{\mathcal{E}}_1(t)\cap\tilde{\mathcal{E}}_3(t) = \emptyset$.}

    We now verify $\mathbb{P}(\mathcal{E}_2) \geq 1 - \delta$.  Firstly, note that for all $t \geq T_{\textnormal{burn-in}}$,
    $
        t \geq \frac{2}{(1 - \ln(2))  p_{\textnormal{PE}}} ( \ln ( \frac{\pi^2 (t - T_{\textnormal{burn-in}} + 1)^2 }{6\delta} ) + d \ln  (1 + \frac{16 \sum_{i=1}^t \overline{z}^2 (i, \delta, x_0 ) }{ {c_{\textnormal{PE}}} p_{\textnormal{PE}} (t-1) } ) ) + 1
    $
    holds by \eqref{sysid:eqn:theorem-rls-error-exciting-T-burn-in}, which implies
    \begin{align}
        &e^{-\frac{1}{2}( 1 - \ln(2) ) p_{\text{PE}}(t-1)} \\
        &\leq  \frac{6\delta}{\pi^2 (t - T_{\textnormal{burn-in}} + 1)^2 } \mleft (1 + \frac{16 \sum_{i=1}^t \overline{z}^2 \mleft (i, \delta, x_0 \mright ) }{ {c_{\textnormal{PE}}} p_{\textnormal{PE}} (t-1) } \mright )^{-d} \\
        &\leq \mathrm{card}\mleft (\mathcal{C}\mleft(t \mright) \mright )^{-1} \frac{6\delta}{\pi^2 (t - T_{\textnormal{burn-in}} + 1)^2 }, \label{sysid:eqn:lemma-regional-pe-6}
    \end{align}
    where \eqref{sysid:eqn:lemma-regional-pe-6} follows from $\mathrm{card}(\mathcal{C}(t)) \leq (1 + 2/\epsilon(t))^d$ and \eqref{sysid:eqn:lemma-regional-pe-1} {if $\sum_{i=1}^t \overline{z}^2(i,\delta,x_0) > 0$, and if $\sum_{i=1}^t \overline{z}^2(i,\delta,x_0) = 0$ it holds via substitution and $\mathrm{card}(\mathcal{C}(t))=1$.}
    Thus, we find that
    \begin{align}
        &\mathbb{P} ( \mathcal{E}_2 ) = {\mathbb{P} \bigg (}  \bigcap_{t \geq T_{\textnormal{burn-in}}} \bigcap_{v \in \mathcal{C} \mleft (t \mright )} \mathcal{E}_1(t,v) {\bigg )} \\
        & \geq
        1 - \sum_{t \geq T_{\textnormal{burn-in}}}      \mathrm{card}\mleft (\mathcal{C}\mleft(t\mright) \mright ) e^{-\frac{1}{2}({1-\ln(2)}){p_{\textnormal{PE}}}(t-1)} \label{sysid:eqn:lemma-regional-pe-10} \\
        & \geq
        1 - \frac{6\delta}{\pi^2}\sum_{t \geq 1} \frac{1}{t^2} = 1 - \delta. \label{sysid:eqn:lemma-regional-pe-12}
    \end{align}
    where \eqref{sysid:eqn:lemma-regional-pe-10} follows from the union bound and because $\mathbb{P}(\mathcal{E}_1(t,v)) \geq 1 - e^{-\frac{1}{2}({1-\ln(2)}){p_{\textnormal{PE}}}(t-1)}$ for all $t \in \mathbb{N}$ and $v \in \mathcal{S}^{d-1}$,
    {and \eqref{sysid:eqn:lemma-regional-pe-12} follows by \eqref{sysid:eqn:lemma-regional-pe-6} and simplification.}
\qed \end{pf*}

\subsection{Proofs for Example~1} \label{sysid:sec:proofs-example-1}

\begin{pf*}{Proof of Prop.~\ref{sysid:prop:bmsb-failure}}
    Suppose $k \in \mathbb{N}$, $\Gamma_{\textnormal{sb}} \succ 0$ and $p {\in (0,1]}$, {and let $\zeta = [0\ \ 1]^\top$.}
    Let $\gamma_{\textnormal{sbu}} = \sqrt{\zeta^{\top} \Gamma_{\textnormal{sb}} \zeta}$, and note that $\gamma_{\textnormal{sbu}} > 0$.

    Next, let $\Phi(\cdot)$ denote the cumulative distribution function of the standard normal distribution. Let $\{\tilde{x}_i\}_{i=1}^{k{+1}} \subseteq \mathbb{R}$ and $\{\tilde{y}_i\}_{i=1}^{{k}} \subseteq \mathbb{R}$ be sequences such that $\tilde{x}_{{k+1}} = \overline{x}$ holds, and both $1 - \Phi\mleft (\frac{\tilde{x}_{i+1} - (\tilde{y}_i + 1)}{{\sigma_w}} \mright) = \mleft ( 1 - p/2 \mright )^{1/k}$
    and $\tilde{x}_i = \max( \tilde{y}_i, \overline{x} )$ are satisfied for $i = {k},\hdots,1$.
    Note that for all $i \in \{1,\hdots,k\}$, on the event $\{ X(i) > \tilde{x}_i \}$,
    \begin{align}
        &\mathbb{P} \mleft ( X(i+1) > \tilde{x}_{i+1} \mid X(i) \mright ) \\
        &\geq \mathbb{E} \mleft [ \mathbf{1}_{ \mleft \{ 1 + \tilde{y}_i + W(1+i) > \tilde{x}_{i+1} \mright \}} \mid X(i) \mright ] \label{sysid:eqn:bmsb-failure-4} \\
        &\geq \mathbb{P} \mleft ( 1 + \tilde{y}_i + W(1+i) > \tilde{x}_{i+1} \mright ) \label{sysid:eqn:bmsb-failure-5} \\
        &= 1 - \Phi\mleft (\frac{\tilde{x}_{i+1} - (\tilde{y}_i + 1)}{{\sigma_w}} \mright)= \mleft ( 1 - p/2 \mright )^{1/k}, \label{sysid:eqn:bmsbfailure-10}
    \end{align}
    where \eqref{sysid:eqn:bmsb-failure-4} follows from the fact that $1 + X(i) + \mathbf{1}_{\mleft \{  X(i) \leq \overline{x} \mright \}} S(i) + W(i+1) {=} 1 + X(i) + W(i+1)$ and $X(i) > \tilde{y}_i$ on the event $\{X(i) > \tilde{x}_i\}$, and \eqref{sysid:eqn:bmsb-failure-5} follows since $\mathbf{1}_{ \mleft \{ 1 + \tilde{y}_i + W(1+i) > \tilde{x}_{i+1} \mright \}}$ and $X(i)$ are independent.

    Next, we find that for all $i \in \{1,\hdots,k\}$, on the event $\{ X(1) > \tilde{x}_1 \}$,
    \begin{align}
        &\mathbb{P} \mleft (  X(i+1)  > \overline{x} \mid X(1) \mright ) \geq \mathbb{P} \mleft (  X(i+1)  > \tilde{x}_{i+1} \mid X(1) \mright ) \\
        &= \mathbb{E} \mleft[ \mathbf{1}_{\{  X(i+1) > \tilde{x}_{i+1} \}} \mid X(1) \mright] \label{sysid:eqn:bmsb-failure-8} \\
        &\geq
        \mathbb{E} \mleft[ \mathbb{P} \mleft ( X(i+1)  > \tilde{x}_{i+1} \mid X(i) \mright ) \mathbf{1}_{\{ X(i) > \tilde{x}_i \}} \mid X(1) \mright] \quad \label{sysid:eqn:bmsb-failure-7} \\
        &\geq \mleft (1 - p/2 \mright )^{1/k} \mathbb{E} \mleft[ \mathbf{1}_{\{ X(i) > \tilde{x}_i \}} \mid X(1) \mright] \label{sysid:eqn:bmsb-failure-9} \\
        & \geq \mleft ( 1 - p/2 \mright )^{i/k}   \geq 1 - p/2, \label{sysid:eqn:bmsb-failure-1}
    \end{align}
    where \eqref{sysid:eqn:bmsb-failure-7} follows from i) the monotonicity of conditional expectation, ii) the tower property, and iii) the conditional independence of $\mathbf{1}_{\{  X(i+1)  > \tilde{x}_{i+1} \}}$ and $X(1)$ given $X(i)$, \eqref{sysid:eqn:bmsb-failure-9} follows from \eqref{sysid:eqn:bmsbfailure-10}, and \eqref{sysid:eqn:bmsb-failure-1} follows by repeating the steps from \eqref{sysid:eqn:bmsb-failure-8} to \eqref{sysid:eqn:bmsb-failure-9} $i-1$ more times.

    Now, note that for all $i \in \{1, \hdots, k \}$, on the event $  \{ X(i+1)  > \overline{x}  \}$, $ \vert \mathbf{1}_{\{ X(i+1) \leq \overline{x} \}} U(i+1)  \vert = 0 < \gamma_{\textnormal{sbu}}$. From this, we have
    $
        \mathbb{P}  (  X(i+1) > \overline{x} \mid X(1)  )
        \leq
        \mathbb{P}  (  \vert \mathbf{1}_{\{ X(i+1) \leq \overline{x} \}} U(i+1)  \vert < \gamma_{\textnormal{sbu}} \mid X(1) ).
    $
    Combining with \eqref{sysid:eqn:bmsb-failure-1}, for all $i \in \{1,\hdots,k\}$, on the event $\mleft \{ X(1) \geq \tilde{x}_1 \mright \}$,
    \begin{align}
        &\mathbb{P} \mleft ( \mleft \vert \begin{bmatrix} 0 & 1 \end{bmatrix}Z(2+i) \mright \vert \geq \sqrt{ \zeta^{\top} \Gamma_{\textnormal{sb}} \zeta } \mid \mathcal{F}(2) \mright ) \\
        &= 1 -  \mathbb{P} \mleft ( \mleft \vert \mathbf{1}_{\{ X(i+1) \leq \overline{x} \}} U(i+1) \mright \vert < \gamma_{\textnormal{sbu}} \mid X(1) \mright ) \label{sysid:eqn:bmsb-failure-10}\\
        &\leq 1 - \mathbb{P} \mleft (  X(i+1) > \overline{x} \mid X(1) \mright ) \leq p/2 < p, \label{sysid:eqn:bmsb-failure-13}
    \end{align}
    {where \eqref{sysid:eqn:bmsb-failure-10} follows from $Z(2)=\psi(X(1),U(1))=[X(1) \ \mathbf{1}_{ \{ X(1)\leq \overline{x} \} }U(1)]^{\top}=[X(1) \ 0]^{\top}$ on the event $\{X(1)\geq\tilde{x}_1\}$.}
    Finally, by choosing $j=2$ and $\zeta = [0 \ \ 1]^{\top}$, we conclude that
    $
        \mathbb{P}  ( \frac{1}{k} \sum_{i=1}^k \mathbb{P}  (  \vert \zeta^{\top}Z(2+i)  \vert \geq \sqrt{ \zeta^{\top} \Gamma_{\textnormal{sb}} \zeta } \mid \mathcal{F}(j)  ) < p  )
        \geq \mathbb{P}  ( X(1) > \tilde{x}_1  )
        > 0.
    $
    Here, the first inequality follows from \eqref{sysid:eqn:bmsb-failure-13}, and the second inequality follows from the fact that $X(1) = 1 + x_0 + 0.1 \mathbf{1}_{\{  x_0 \leq \overline{x} \}} \cdot S({0}) + W(1)$ has a probability distribution with an unbounded support since $W(1)$ is normally distributed such that $\mathbb{P}( X(1) > \tilde{x}_1 ) > 0$.
\qed \end{pf*}

\begin{pf*}{Proof of Prop.~\ref{sysid:prop:pwa-regional-excitation-success}}
For all $x \in \mathcal{X}_{\textnormal{PE}}$ and $\zeta = (\zeta_1,\zeta_2) \in \mathcal{S}^{1}$,
$
    \mathbf{E}  [  \vert \zeta_1( x + W ) + \zeta_2 \mathbf{1}_{\{  x + W \leq \overline{x} \}} \cdot S  \vert  ]
    \geq
    \mathbf{E}  [ \mathbf{E}  [  \vert \zeta_1( x + W ) + \zeta_2 S  \vert \mid W  ] \cdot \mathbf{1}_{ \{ \vert W  \vert \leq 0.1 \overline{x}  \}}  ]
    \geq
     \vert \zeta_1  \vert \mathbf{E}  [  \vert  x + W  \vert \mid \vert W  \vert \leq 0.1 \overline{x}  ] \mathbf{P} ( \vert W  \vert \leq 0.1 \overline{x}  )
    \geq
    \vert \zeta_1 \vert b_w
$
and moreover,
$\mathbf{E}  [  \vert \zeta_1( x + W ) + \zeta_2 \mathbf{1}_{\{  x + W \leq \overline{x} \}} \cdot S  \vert  ] \geq
    \mathbf{E}  [ \mathbf{E}  [  \vert \zeta_1( x + W ) + \zeta_2 S  \vert \mid S, \mathbf{1}_{ \{ \vert W  \vert \leq 0.1 \overline{x}  \}}  ] \cdot \mathbf{1}_{ \{ \vert W  \vert \leq 0.1 \overline{x}  \}}  ] \geq
    \mathbf{E}  [  \vert \zeta_1 x + \zeta_2 S  \vert   ] \mathbb{P}  ( \vert W  \vert \leq 0.1 \overline{x}  ) \geq
     \vert \zeta_2  \vert b_s $.
Together, these imply that
$
    \mathbf{E}  [  \vert \zeta_1( x + W ) + \zeta_2 \mathbf{1}_{\{  x + W \leq \overline{x} \}} \cdot S  \vert  ]
    \geq \max  ( \vert \zeta_1  \vert b_w , \vert \zeta_2 \vert b_s  ) \geq \frac{b_w b_s}{\sqrt {b_w^2 + b_s^2 }} =c_{\textnormal{PE1}}
$
where we used the fact that $\max  ( \vert \zeta_1  \vert b_w , \vert \zeta_2 \vert b_s  ) $ is minimised when $\vert \zeta_1 \vert = \sqrt{\frac{b_s^2}{b_w^2 + b_s^2}}$ and $\vert \zeta_2 \vert = \sqrt{\frac{b_w^2}{b_w^2 + b_s^2}}$ (since they satisfy $\vert \zeta_1 \vert b_w = \vert \zeta_2 \vert b_s$ and $ \zeta_1^2 + \zeta_2^2 = 1$).

Moreover, for all $x \in \mathcal{X}_{\textnormal{PE}}$, we have
$
    \mathbf{V}\mathrm{ar}  (  \vert
        \zeta_1 (x + W) +
        \zeta_2 (\mathbf{1}_{\{  x + W \leq \overline{x} \}} \cdot S)
    \vert  )
    \leq
    \zeta_1^2 \sigma_w^2 + \zeta_2^2 \frac{1}{3}\overline{u}^2 \leq \max(\sigma_w^2,\frac{1}{3}\overline{u}^2) = c_{\textnormal{PE2}}
$.
{Finally, we establish that $(\psi,\alpha,\mu_s,\mu_w)$ is $(\mathcal{X}_{\textnormal{PE}},c_{\textnormal{PE}},p_{\textnormal{PE}})$\textit{-regionally excited} using Lem.~\ref{sysid:lemma:regional-excitation-sufficient}.}
\qed \end{pf*}
\section{Conclusion} \label{sysid:sec:sys-id-conclusion}

We {studied} the {RLS} identification problem for unstable, stochastic, nonlinear systems with linearly parametrised uncertainty, under arbitrary policies that may not be stabilising.
{To this end, we established sub-exponential growth of the Gramian, primarily relying on a sub-exponential input-to-state growth property on the open-loop dynamics. RLS error bounds that hold for some initial states, failure probabilities, and time intervals were then derived under an additional relaxed PE assumption. Regional excitation was introduced as a condition for certifying relaxed PE, and} characterises the usefulness of a state space region for producing informative regressor data.
This was demonstrated on a PWA system example satisfying regional excitation but not other similar conditions in the literature.
We showed that by {assuming} global excitation instead, the bounds hold over all time.
This was exemplified by a double integrator controlled by a not-necessarily-stabilising policy.

There are several interesting extensions to consider. In the control literature, continuity-type assumptions are required to establish robustness to perturbations \cite{grammatico2013discrete}. We did not study the robustness of the estimates to such perturbations or to nonparametric errors, although these issues are of practical importance.
Another relevant {extension is to consider settings in which Def.~\ref{sysid:def:excitation} fails because the exploratory inputs and process noise do not directly excite every direction of the regressor space, but PE in the sense of Ass.~\ref{assump:interval-pe} may hold due to the propagation of signals through the dynamics. Such settings could potentially be handled by extending Def.~\ref{sysid:def:excitation} to allow $k>1$, analogously to the linear-system analysis in \cite{tsiamis2021linear}.}
{Lastly, the investigation of} correlated exploratory inputs {and process noise is of interest}.
This could involve showing that {the chosen controls and noise} conditions lead to PE (e.g., \cite{mania2020active} does this {using active learning} with intentionally chosen controls that may be correlated), or more broadly, $\ln(\lambda_{\textnormal{max}}(G(t)))/\lambda_{\textnormal{min}}(G(t)) \rightarrow 0$.

\section{Declaration of generative AI and AI-assisted technologies in the manuscript preparation process}
{During the revision of this work, the authors used OpenAI Codex (GPT-5.6 Sol) to assist with checking mathematical derivations, coding and improving language.
After using this tool/service, the authors reviewed and edited the content as needed and take full responsibility for the content of the published article.
}

\bibliographystyle{plain}

\ifarxivversion
    \appendix
\section{Technical Lemmas} \label{sysid:sec:technical-lemmas}

The proofs of the following technical lemmas are provided in Sec.~\ref{sysid:sec:proof-technical-lemmas}.

\begin{lemma} \label{sysid:lemma:sub-gaussian-uniform-bound}
    (Bound on sub-Gaussian sequence uniformly over time)
    Consider {an $\mathbb{R}^d$-valued} random sequence $\{Y(t)\}_{t \in \mathbb{N}}$ such that $Y(t)$ is $\sigma_y^2$-sub-Gaussian for all $t \in \mathbb{N}$. Then, for any $\delta \in (0,1)$,
    \begin{align}
        P \mleft ( \mleft \vert Y(t) \mright \vert \leq \sigma_y \sqrt{ 2 {d} \ln \mleft( \frac{{d}\pi^2t^2}{3\delta} \mright )  }, \ \forall t \in \mathbb{N} \mright ) \geq 1 - \delta.
    \end{align}
\end{lemma}

\begin{lemma} \label{sysid:lemma:joint-chernoff-bound} (Joint Chernoff bound)
    Consider non-negative random variable $X$, and an event $\mathcal{E}$. Suppose that $\mathbb{E}[\bm{1}_{\mathcal{E}} e^{sX}] < \infty$ for any $s < 0$. Then, for all $s < 0$ and $a > 0$,
    \begin{align}
        P( \mathcal{E} \cap \{ X \leq a \}) \leq \mathbb{E}[\bm{1}_{\mathcal{E}} e^{sX}] e^{-sa}.
    \end{align}
\end{lemma}

\begin{lemma} (Joint Markov inequality) \label{sysid:lemma:joint-markov-bound}
    Consider a non-negative scalar random variable $X$ and an event $\mathcal{E}$. Then, for all $a > 0$,
    \begin{align}
        P(\mathcal{E} \cap \{X \geq a\}) \leq \mathbb{E}[\bm{1}_{\mathcal{E}} X]/a.
    \end{align}
\end{lemma}

\begin{lemma} \label{sysid:lemma:simplify-min-eigenvalue-bound}
    For all $p \in (0,1)$,
    \begin{align}
         \inf_{s<0} p e^{s(1-\frac{p}{2})} + (1-p)e^{-s \frac{p}{2}}  \leq e^{-\frac{1}{2} ({1 - \ln(2)})p}. \label{sysid:eqn:lemma-simplify-min-eigenvalue-bound-1}
    \end{align}
\end{lemma}

\begin{lemma} \label{sysid:lemma:sub-exp-equiv}
    Consider {an eventually positive} function $g:\mathbb{R}_{\geq 0} \rightarrow \mathbb{R}_{\geq 0}$. If $ \ln (g(r)) = o(r)$ then $g ( \ln (r) ) = o(r)$.
\end{lemma}

\begin{lemma} \label{sysid:lemma:sub-exp-sum}
    Consider {the eventually positive functions} $g,h:\mathbb{R}_{\geq 0} \rightarrow \mathbb{R}_{\geq 0}$. If $\ln ( g (r) ) = o(r)$ and $\ln ( h ( r) ) = o(r)$, then $\ln ( g(r) + h(r) ) = o(r)$.
\end{lemma}

\begin{lemma} \label{sysid:lemma:sub-exp-constant}
    Consider {the eventually positive function} $g:\mathbb{R}_{\geq 0} \rightarrow \mathbb{R}_{\geq 0}$. If $\ln ( g (r) ) = o(r)$ then for all $c > 0$, $\ln ( g( c r) ) = o(r)$.
\end{lemma}
\section{Supplementary Results} \label{sysid:sec:supplementary-results}

\subsection{Proof of Regional Excitation via Moments} \label{sysid:sec:proof-regional-excitation-moments}

\begin{pf*}{Proof of Lem.~\ref{sysid:lemma:regional-excitation-sufficient}}
    {Suppose $\vartheta \in \mathbb{R}^{d \times n}$, $x \in \mathcal{X}$ and $\zeta \in \mathcal{S}^{d-1}$. Then,
    \begin{align}
        &\mathbf{P}\mleft (  \mleft \vert \zeta^{\top} \psi \mleft ( x + W , \alpha(x+W,S,\vartheta) \mright ) \mright \vert^2 \geq c_{\textnormal{PE}} \mright ) \\
        &=\mathbf{P} \mleft ( \mleft \vert \zeta^{\top} \psi \mleft ( x + W , \alpha(x+W,S,\vartheta) \mright ) \mright \vert \geq \frac{1}{2} c_{\textnormal{PE}1} \mright ) \\
        &\geq \mathbf{P} \Big( \mleft \vert \zeta^{\top} \psi \mleft ( x + W , \alpha(x+W,S,\vartheta) \mright ) \mright \vert \geq \\
        & \quad \frac{1}{2} \mathbf{E} \mleft [ \mleft \vert \zeta^{\top} \psi \mleft ( x + W , \alpha(x+W,S,\vartheta) \mright ) \mright \vert  \mright ]  \Big) \\
        &\geq \frac{1}{4} \mleft ( \frac{\mathbf{V}\mathrm{ar}\mleft (\mleft \vert \zeta^{\top} \psi \mleft ( x + W , \alpha(x+W,S,\vartheta) \mright ) \mright \vert \mright )}{\mathbf{E} \mleft [ \mleft \vert \zeta^{\top} \psi \mleft ( x + W , \alpha(x+W,S,\vartheta) \mright ) \mright \vert \mright ]^2} + 1 \mright )^{-1} \\
        & \geq \frac{1}{4} \mleft ( \frac{ c_{\textnormal{PE}2}}{c_{\textnormal{PE}1}^2} + 1 \mright )^{-1} = p_{\textnormal{PE}},
    \end{align}
    where the second-last inequality follows from the Paley-Zygmund inequality \cite[Lem.~2.86]{zygmund2002trigonometric}.}
\qed \end{pf*}

\subsection{Proof of Data-Dependent Least Squares Error Bound} \label{sysid:sec:proof-data-dependent-error-bound}

{We prove Lem.~\ref{sysid:lemma:data-dependent-error-bound} in this section. To do so, we make use of a standard result on self-normalised bounds for vector-valued martingales in Lem.~\ref{sysid:lemma:self-normalised-bound}.}
\begin{lemma} (Self-Normalised Bound for Vector-Valued Martingales \cite[Thm.~1]{abbasi2011improved}) \label{sysid:lemma:self-normalised-bound}
    \\Let $\{ \mathcal{G}(t) \}_{t \in \mathbb{N}_0} \subseteq \mathcal{F}$ be a filtration. Let $\{ \eta(t) \}_{t \in \mathbb{N}}$ be any real-valued random sequence such that $\eta(t)$ is $\mathcal{G}(t)$-measurable and $\eta(t) \mid \mathcal{G}(t-1)$ is conditionally $\sigma_{\eta}^2$-sub-Gaussian for some $\sigma_{\eta} \geq 0$. Let $\{Z(t)\}_{t \in \mathbb{N}}$ be any $\mathbb{R}^d$-valued random sequence such that $Z(t)$ is $\mathcal{G}(t-1)$-measurable. Let $R(t) = \sum_{i=1}^t \eta(i) Z(i) $ and $G(t) = \sum_{i = 1}^t Z(i) Z(i)^{\top} + v$ for $t \in \mathbb{N}_0$, where $v$ is some $d \times d$ positive definite matrix. Then, for any $\delta_1 > 0$,
    \begin{align}
        &P \mleft( \mleft \vert R(t) \mright \vert_{G(t)^{-1}}^2 \leq 2 \sigma_{\eta}^2 \ln \mleft( \frac{\det \mleft( G(t) \mright)^{1/2} \det \mleft( v \mright)^{-1/2}}{\delta_1} \mright), \mright. \\
        & \quad \mleft.  \forall t \in \mathbb{N} \mright) \geq 1 - \delta_1.
    \end{align}
\end{lemma}

\begin{pf*}{Proof of Lem.~\ref{sysid:lemma:data-dependent-error-bound}}
    {For all $t \in \mathbb{N}$, the least squares estimate can be rewritten as
    \begin{align}
        &\hat{\theta}(t) = \argmin_{\theta \in \mathbb{R}^{d \times n}} \sum_{s=1}^t \vert X(s) - f(X(s-1),U(s-1)) \\
        &\quad - \theta^{\top} Z(s) \vert^2 + \gamma \vert \theta \vert_F^2 \\
        &= G(t)^{-1} \sum_{s=1}^t Z(s)W(s)^{\top}  + \theta_* - \gamma G(t)^{-1} \theta_*. \label{sysid:eqn:lemma-rls-interface-1}
    \end{align}
    By rearranging \eqref{sysid:eqn:lemma-rls-interface-1}, we have
    \begin{align}
        &G(t)^{1/2} \mleft (\hat{\theta}(t) - \theta_* \mright ) \\
        &= G(t)^{-1/2} \sum_{s=1}^t Z(s)W(s)^{\top}  - \gamma G(t)^{-1/2} \theta_*. \label{sysid:eqn:lemma-rls-interface-2}
    \end{align}
    Then, by taking the {$2$-}norm of both sides of \eqref{sysid:eqn:lemma-rls-interface-2}, we have
    \begin{align}
        &\mleft \vert G(t)^{1/2} \mleft (\hat{\theta}(t) - \theta_* \mright ) \mright \vert \\
        &\leq \mleft \vert G(t)^{-1/2} \sum_{s=1}^t Z(s)W(s)^{\top}  \mright \vert + \gamma \mleft \vert G(t)^{-1/2} \theta_* \mright \vert \label{sysid:eqn:lemma-rls-interface-3} \\
        &\leq  \mleft \vert G(t)^{-1/2} \sum_{s=1}^t Z(s)W(s)^{\top} \mright \vert + \gamma^{1/2}\mleft \vert \theta_* \mright \vert \label{sysid:eqn:lemma-rls-interface-main}
    \end{align}
    almost surely. Here, \eqref{sysid:eqn:lemma-rls-interface-3} follows from the triangle inequality, and \eqref{sysid:eqn:lemma-rls-interface-main} follows since
    {$ \lvert G(t)^{-1/2} \theta_* \rvert \leq \lvert G(t)^{-1/2} \rvert \lvert \theta_* \rvert \leq \gamma^{-1/2}\lvert \theta_* \rvert $}
    almost surely, substituting these inequalities into \eqref{sysid:eqn:lemma-rls-interface-3}, then simplifying.}

    Next, from the definition of the {2}-norm, we have
    \begin{align}
        \mleft \vert G(t)^{-1/2}  \sum_{i=1}^t Z(i) W(i)^{\top}  \mright \vert^2 = {\sup_{v \in \mathcal{S}^{n-1}}} \mleft \vert \sum_{i=1}^t Z(i) W(i)^{\top} {v} \mright \vert_{G(t)^{-1}}^2 \label{sysid:eqn:lemma-self-normalised-bound-main-1}
    \end{align}
    almost surely.
     Next, let $\{\mathcal{G}(t)\}_{t \in \mathbb{N}_0} \subseteq \mathcal{F}$ be the filtration satisfying $\mathcal{G}(t) = \sigma( \{ Z(i) \}_{i = 1}^{t+1}, \{ W(i) \}_{i = 1}^t )$.
     Since {for all $v \in \mathcal{S}^{n-1}$, $W(t)^{\top} v$} is $\mathcal{G}(t)$-measurable, {$W(t)^{\top} v \mid \mathcal{G}(t-1)$ is conditionally} $\sigma_w^2$-sub-Gaussian {due to Ass.~\ref{sysid:assump:process-noise},} and $Z(t)$ is $\mathcal{G}(t-1)$-measurable, from Lem.~\ref{sysid:lemma:self-normalised-bound} we have that for any {$v \in \mathcal{S}^{n-1}$} and $\delta \in (0,1)$,
    \begin{align}
        &P\mleft (\mleft \vert G(t)^{-1/2} \sum_{i=1}^t Z(i) {W(i)^{\top} v} \mright \vert^2 \leq \mright. \\
        &\mleft. 2 \sigma_w^2 \ln \mleft( \frac{\det \mleft( G(t) \mright)^{1/2} \det \mleft( \gamma I \mright)^{-1/2}}{\delta} \mright), \ \forall t \in \mathbb{N} \mright ) \\
        &\geq 1 - \delta. \label{sysid:eqn:lemma-self-normalised-bound-main-2}
    \end{align}
    {Let $\epsilon=1/2$, and $\mathcal{C}$ be an $\epsilon$-cover of $\mathcal{S}^{n-1}$ with cardinality at most $5^n$ (verified via \cite[Cor.~4.2.13]{vershynin2018high}). It follows that
    $\lvert G(t)^{-1/2} \sum_{i=1}^t Z(i)W(i)^{\top} \rvert^2 \leq \frac{1}{(1-\epsilon)^2} \sup_{v \in \mathcal{C}} \lvert G(t)^{-1/2} \sum_{i=1}^t Z(i)W(i)^{\top} v \rvert^2 = 4 \sup_{v \in \mathcal{C}} \lvert G(t)^{-1/2} \sum_{i=1}^t Z(i)W(i)^{\top} v \rvert^2$,
    which combined with \eqref{sysid:eqn:lemma-self-normalised-bound-main-2} and a union bound argument {over $v \in \mathcal{C}$ with failure probability $\delta/5^n$ at each point} implies
    \begin{align}
        &P\mleft ( \mleft \vert G(t)^{-1/2} \sum_{i=1}^t Z(i) W(i)^{\top} \mright \vert^2 \leq \mright. \\
        &\mleft. 8 \sigma_w^2 \ln \mleft( \frac{5^n \det \mleft( G(t) \mright)^{1/2} \det \mleft( \gamma I \mright)^{-1/2}}{\delta} \mright), \ \forall t \in \mathbb{N} \mright ) \\
        & \geq 1 - \delta, \label{sysid:eqn:lemma-self-normalised-bound-main-3}
    \end{align}
    for any $\delta \in (0,1)$.}

    We arrive at the conclusion by combining \eqref{sysid:eqn:lemma-rls-interface-main}, \eqref{sysid:eqn:lemma-self-normalised-bound-main-1}, and \eqref{sysid:eqn:lemma-self-normalised-bound-main-3}, and noting that
    $\mleft \vert G(t)^{1/2} \mleft ( \hat{\theta}(t) - \theta_* \mright) \mright \vert \geq \lambda_{\textnormal{min}}(G(t))^{1/2} \mleft \vert  \hat{\theta}(t) - \theta_*  \mright \vert $
    and \\$\ln \mleft( \frac{{5^n} \det \mleft( G(t) \mright)^{1/2} \det \mleft( \gamma I \mright)^{-1/2}}{\delta} \mright) \leq \ln (1 / \delta) + {n \ln (5) + } (d/2) \ln \mleft( \lambda_{\textnormal{max}}(G(t)) \gamma^{-1} \mright)$ both hold.
\qed \end{pf*}

\subsection{Proof of Technical Lemmas} \label{sysid:sec:proof-technical-lemmas}

\begin{pf*}{Proof of Lem.~\ref{sysid:lemma:sub-gaussian-uniform-bound}}
    {Let $Y_j(t)$ denote the $j$th coordinate of $Y(t)$. }
    Using the standard Chernoff bound result for sub-Gaussian random variables, we find that for all $t \in \mathbb{N}$, {$j\leq d$} and $a > 0$,
    \begin{align}
        P \mleft ( \mleft \vert Y_{{j}}(t) \mright \vert > a \mright ) \leq 2 e^{- \frac{a^2}{2 \sigma_y^2}}. \label{sysid:eqn:lemma-sub-gaussian-uniform-bound-1}
    \end{align}

    Suppose $\delta \in (0,1)$.
    Substituting $a$ in \eqref{sysid:eqn:lemma-sub-gaussian-uniform-bound-1} for $\sigma_y \sqrt{ 2  \ln \mleft( \frac{{d}\pi^2t^2}{3\delta} \mright ) }$, we have for all $t \in \mathbb{N}$ {and $j \leq d$},
    \begin{align}
        P \mleft ( \mleft \vert Y_{{j}}(t) \mright \vert > \sigma_y \sqrt{ 2  \ln \mleft( \frac{{d}\pi^2t^2}{3\delta} \mright ) } \mright ) \leq \frac{6 \delta}{{d}\pi^2 t^2}
    \end{align}
    Then, using a union bound argument:
    \begin{align}
        &P \mleft ( \mleft \vert Y(t) \mright \vert \leq \sigma_y \sqrt{ 2 d  \ln \mleft( \frac{d \pi^2t^2}{3\delta} \mright ) }, \ \forall t \in \mathbb{N} \mright ) \\
        &\geq P \mleft ( \mleft \vert Y_j(t) \mright \vert \leq \sigma_y \sqrt{ 2  \ln \mleft( \frac{d \pi^2t^2}{3\delta} \mright ) }, \ \forall t \in \mathbb{N}, j \leq d \mright ) \\
        &\geq 1- \sum_{j=1}^d \frac{6 \delta}{d \pi^2} \sum_{t \in \mathbb{N}} \frac{1}{t^2} = 1- \frac{6 \delta}{\pi^2} \sum_{t \in \mathbb{N}} \frac{1}{t^2} = 1 - \delta.
    \end{align}
    This concludes the proof. \qed
\end{pf*}

\begin{pf*}{Proof of Lem.~\ref{sysid:lemma:joint-chernoff-bound}}
    Suppose $s < 0$ and $a > 0$. Then, since $s$ is negative,
    \begin{align}
        P( \mathcal{E} \cap \{X \leq a\}) = P( \mathcal{E} \cap \{ e^{sX} \geq e^{sa} \}) \leq \frac{ \mathbb{E}[\bm{1}_{\mathcal{E}} e^{sX}]}{e^{sa}},
    \end{align}
    where the inequality follows from Lem. \ref{sysid:lemma:joint-markov-bound}. This concludes the proof.
\qed \end{pf*}

\begin{pf*}{Proof of Lem.~\ref{sysid:lemma:joint-markov-bound}}
    Suppose $a > 0$. The result holds by rearranging the following inequality:
    \begin{align}
        &\mathbb{E}[\bm{1}_{\mathcal{E}} X] = \mathbb{E}[ X \bm{1}_{\mathcal{E}} \bm{1}_{\{X < a\}} ] + \mathbb{E}[ X \bm{1}_{\mathcal{E}} \bm{1}_{\{X \geq a\}} ] \\
        &\geq \mathbb{E}[ X \bm{1}_{\mathcal{E}} \bm{1}_{\{ X \geq a \}}  ] \geq \mathbb{E}[ a \bm{1}_{\mathcal{E}} \bm{1}_{\{ X \geq a \}}  ] \\
        &= a P( \mathcal{E} \cap \{ X \geq a \} ).
    \end{align}
\qed \end{pf*}

\begin{pf*}{Proof of Lem.~\ref{sysid:lemma:simplify-min-eigenvalue-bound}}
    Differentiating the left hand side of \eqref{sysid:eqn:lemma-simplify-min-eigenvalue-bound-1} with respect to $s$, we find:
    \begin{align}
        &\frac{d}{ds} p e^{s(1-\frac{p}{2})} + (1-p)e^{-s \frac{p}{2}} \\
        &= pe^{s(1-\frac{p}{2})}(1- \frac{p}{2}) + (1 - p)e^{-s(\frac{p}{2})}(-\frac{p}{2}). \label{sysid:eqn:lemma-simplify-min-eigenvalue-bound-2}
    \end{align}
    Since the left hand side of \eqref{sysid:eqn:lemma-simplify-min-eigenvalue-bound-2} is convex in $s$, we can solve for the optimum by setting \eqref{sysid:eqn:lemma-simplify-min-eigenvalue-bound-2} to zero and solving for $s$, yielding $s_* = \ln ( \frac{1 - p}{2 - p} )$. Substituting $s_*$ back into the left hand side of \eqref{sysid:eqn:lemma-simplify-min-eigenvalue-bound-1}, we find
    \begin{align}
        &\inf_{s<0} p e^{s(1-\frac{p}{2})} + (1-p)e^{-s \frac{p}{2}} \\
        &= p e^{s_*(1-\frac{p}{2})} + (1-p)e^{-s_* \frac{p}{2}} \\
        &=p e^{\ln ( \frac{1 - p}{2 - p} )(1-\frac{p}{2})} + (1-p)e^{-\ln ( \frac{1 - p}{2 - p} ) \frac{p}{2}} \\
        &= 2 \mleft( \frac{1 - p}{2 - p} \mright)^{1 - \frac{p}{2}} \\
        &=  \exp \mleft( \ln \mleft( 2 ( \frac{1 - p}{2 - p} )^{1 - \frac{p}{2}} \mright) \mright) \label{sysid:eqn:lemma-simplify-min-eigenvalue-bound-3}
    \end{align}

    Next we derive a simplified upper bound. The coefficients of the Taylor series expansion of $\ln ( 2 ( \frac{1 - p}{2 - p} )^{1 - \frac{p}{2}} )$ around zero are all non-positive, and therefore
    \begin{align}
        &\ln \mleft( 2 \mleft( \frac{1 - p}{2 - p} \mright)^{1 - \frac{p}{2}} \mright) \leq \frac{d}{d p} \ln \mleft( 2 \mleft( \frac{1 - p}{2 - p} \mright)^{1 - \frac{p}{2}} \mright) \bigg \vert_{p=0} p \\
        &= -\frac{1}{2} (1 - \ln(2))p \label{sysid:eqn:lemma-simplify-min-eigenvalue-bound-4}
    \end{align}

    The conclusion follows by combining \eqref{sysid:eqn:lemma-simplify-min-eigenvalue-bound-3}  with \eqref{sysid:eqn:lemma-simplify-min-eigenvalue-bound-4}.
\qed \end{pf*}

\begin{pf*}{Proof of Lem.~\ref{sysid:lemma:sub-exp-equiv}}
    Suppose $\ln ( g (r) ) = o(r)$, such that for all $\epsilon_1 > 0$, there exists $\delta_1 > 0$ such that for all $s > \delta_1$,
    \begin{align}
        \frac{\ln( g (s))}{s} < \epsilon_1,
    \end{align}
    which is equivalent to
    \begin{align}
        g(s) < \exp ( s \epsilon_1 ). \label{sysid:eqn:lemma-sub-exp-equiv-1}
    \end{align}
    It follows that for all $\epsilon_1 > 0$, there exists $\delta_1>0$ (satisfying \eqref{sysid:eqn:lemma-sub-exp-equiv-1}) such that for all $r$ satisfying $r > \exp (\delta_1 )$,
    \begin{align}
        g ( \ln ( r ) ) < \exp ( \ln ( r ) \epsilon_1 ) = r^{\epsilon_1}. \label{sysid:eqn:lemma-sub-exp-equiv-2}
    \end{align}

    Suppose $\epsilon_2 > 0$. Let $\epsilon_1$ satisfy $\epsilon_1 \in (0,1)$, and let $\delta_1$ satisfy \eqref{sysid:eqn:lemma-sub-exp-equiv-2}.
    Let $\delta_2 > 0$ be such that for all $r > \delta_2$,
    \begin{align}
        r^{\epsilon_1 - 1} < \epsilon_2,
    \end{align}
    where satisfactory $\delta_2$ exists since $r^{\epsilon_1 - 1} \rightarrow 0$ as $r \rightarrow \infty$. It follows that if $r > \max ( \exp (\delta_1), \delta_2 )$,
    \begin{align}
        \frac{g( \ln ( r ) )}{r} \leq \frac{r^{\epsilon_1}}{r} = r^{\epsilon_1 - 1} < \epsilon_2.
    \end{align}
    {Lastly, since $g$ is non-negative, we have $g(\ln r)/r\geq0$ for all $r \geq 1$.}
    This concludes the proof.
\qed \end{pf*}

\begin{pf*}{Proof of Lem.~\ref{sysid:lemma:sub-exp-sum}}
    Suppose $\ln ( g (r) ) = o(r)$ and $\ln ( h ( r) ) = o(r)$. This means that for all $\epsilon_1 > 0$, there exists $\delta_1 > 0$ such that for all $r > \delta_1$,
    \begin{align}
        \frac{\ln ( g ( r) )}{r} < \epsilon_1 \quad \text{and} \quad \frac{\ln ( h ( r) )}{r} < \epsilon_1.
    \end{align}
    which implies
    \begin{align}
        g(r) + h(r) < 2 \exp ( r \epsilon_1),
    \end{align}
    which is equivalent to
    \begin{align}
        \frac{\ln ( g(r) + h(r) )}{r} < \epsilon_1 + \frac{\ln(2)}{r}. \label{sysid:eqn:lemma-sub-exp-sum-1}
    \end{align}

    Now, suppose $\epsilon_2 > 0$, let $\epsilon_1$ satisfy $\epsilon_1 = \frac{\epsilon_2}{2}$, let $\delta_1$ satisfy \eqref{sysid:eqn:lemma-sub-exp-sum-1}, and let $\delta_2$ satisfy $\delta_2 = \frac{\ln(2)}{\epsilon_2 - \epsilon_1}$. It follows that if $r > \max(\delta_1,\delta_2)$,
    \begin{align}
        \epsilon_1 + \frac{\ln(2)}{r} < \epsilon_2, \label{sysid:eqn:lemma-sub-exp-sum-2}
    \end{align}
    The conclusion follows after combining \eqref{sysid:eqn:lemma-sub-exp-sum-1} with \eqref{sysid:eqn:lemma-sub-exp-sum-2}, {and by noting that $\liminf_{r \rightarrow \infty}\ln(g(r)+h(r))/r \geq \liminf_{r \rightarrow \infty}\ln(g(r))/r=0$}.
\qed \end{pf*}

\begin{pf*}{Proof of Lem.~\ref{sysid:lemma:sub-exp-constant}}
    Suppose $\ln ( g (r) ) = o(r)$, such that
    \begin{align}
        \lim_{r \rightarrow \infty} \frac{\ln ( g ( r ) )}{r} = 0. \label{sysid:eqn:lemmasub-exp-constant-1}
    \end{align}

    Suppose $c > 0$. Then, we have
    \begin{align}
        \lim_{r \rightarrow \infty} \frac{\ln ( g (cr))}{r} = c \lim_{s \rightarrow \infty} \frac{\ln ( g (s))}{s} = c \cdot 0 = 0,
    \end{align}
    where we apply the change of variable $r = \frac{s}{c}$ and make use of \eqref{sysid:eqn:lemmasub-exp-constant-1}.
\qed \end{pf*}
 \fi

\end{document}